\documentclass[english]{elsarticle}
\usepackage[T1]{fontenc}
\usepackage[latin9]{inputenc}
\usepackage{color}
\usepackage{float}
\usepackage{textcomp}
\usepackage{amsmath}
\usepackage{amssymb}
\usepackage{graphicx}

\makeatletter

\providecommand{\tabularnewline}{\\}

\usepackage{fullpage}
\usepackage{url}
\usepackage{xcolor}

\definecolor{green}{RGB}{0, 140, 0}

\makeatother

\usepackage{babel}
\begin{document}
\title{Semi-Lagrangian Vlasov simulation on GPUs\tnoteref{label1}}
\author[uibk]{Lukas Einkemmer\corref{cor1}} \ead{lukas.einkemmer@uibk.ac.at}
\address[uibk]{University of Innsbruck, Austria}
\cortext[cor1]{Corresponding author}
\begin{abstract}
In this paper, our goal is to efficiently solve the Vlasov equation on GPUs. A semi-Lagrangian discontinuous Galerkin scheme is used for the discretization. Such kinetic computations are extremely expensive due to the high-dimensional phase space. The SLDG code, \textcolor{black}{which is publicly available under the MIT license}, abstracts the number of dimensions and uses a shared codebase for both GPU and CPU based simulations. We investigate the performance of the implementation on a range of both Tesla (V100, Titan V, K80) and consumer (GTX 1080 Ti) GPUs. Our implementation is typically able to achieve a performance of approximately 470 GB/s on a single GPU and 1600 GB/s on four V100 GPUs connected via NVLink. This results in a speedup of about a factor of ten (comparing a single GPU with a dual socket Intel Xeon Gold node) and approximately a factor of 35 (comparing a single node with and without GPUs). In addition, we \textcolor{black}{investigate} the effect of single precision computation on the performance of the SLDG code and demonstrate that a template based dimension independent implementation can achieve good performance regardless of the dimensionality of the problem.
\end{abstract}
\maketitle

\section{Introduction}

Being able to efficiently run large scale physics simulation on graphic
processing units (GPUs) is increasingly important in a range of applications.
Since frequency scaling essentially ended approximately a decade ago,
almost all performance improvement in computer hardware that has been
achieved since then can be traced back to increased parallelism. This
is certainly true for traditional central processing unit (CPU) based
systems, where both explicit (i.e.~controlled by the programmer)
parallelism, such as multi-core systems and vectorization, as well
as implicit (i.e.~controlled by the hardware) parallelism, such as
instruction level parallelism, are now required to obtain good performance.
GPUs, which have a long history in graphics applications, take the
parallel paradigm even further. There, the frequency of each execution
engine is significantly lower; i.e.~a sequential program on the GPU
would be significantly slower than a comparable program on the CPU.
Since power dissipation scales as the \textcolor{black}{cube} of the
frequency, this enables the hardware to expose even more parallelism
\textcolor{black}{(see, e.g., \cite{horowitz2005,miyoshi2002} for
more details).} Furthermore, modern GPUs are equipped with high bandwidth
memory. This is particularly important for many scientific computing
applications were memory bandwidth is the main driver of performance.
However, obtaining good performance requires that algorithms are developed
and efficiently implemented that work within the constraints of the
hardware (e.g.~regular and predictable access of memory).

GPUs are also becoming more and more important in the race to exascale
computing. In fact, many pre-exascale systems, such as SUMMIT summit,
are based on GPUs. Moreover, it is generally agreed upon that GPUs,
or other accelerators of some sort, will be required to build an exascale
system that operates within a reasonable power envelope. In this setting,
the favorable flop/watt ratio of GPUs is a significant asset. It is
thus no coincidence that GPU based systems dominate the Green 500
ranking\footnote{\url{https://www.top500.org/green500/lists/2019/06/}}
(a list of the most power efficient supercomputers). The fat node
approach, i.e.~multiple GPUs on a single node and thus reducing the
overall number of nodes in a system, seems to be a promising architecture
going forward (for example, SUMMIT is one system that follows this
approach). This also enables the use of fast interconnect, such as
NVIDIA's NVLink, on each node.

In the present paper our goal is to demonstrate the viability of using
GPUs for efficiently performing semi-Lagrangian Vlasov simulations
and to investigate the comparative performance of such an approach.
More specifically, we consider the following partial differential
equation
\[
\partial_{t}f(t,x,v)+v\cdot\nabla_{x}f(t,x,v)-(E(f)(t,x)+v\times B(f)(t,x))\cdot\nabla_{v}f(t,x,v)=C(f),
\]
where $f(t,x,v)$ is the sought after particle-density function. The
electric $E$ and magnetic field $B$ are determined using an appropriate
model of electromagnetic radiation. In the simplest case, i.e.~ for
the Vlasov--Poisson equation, we have $B(f)=0$ (no magnetic effects),
$C(f)=0$ (collisionless plasma), and the electric field is determined
by solving the following Poisson problem
\[
E=-\nabla\phi,\qquad-\Delta\phi=\rho+1,\qquad\rho(f)(t,x)=-\int f(t,x,v)\,\mathrm{dv}.
\]

This is the model we will primarily consider in the present paper.
However, let us note that the algorithms used and the framework described
in section \ref{sec:algorithms-and-code} are also able to handle
configurations with $B(f)\neq0$ and $C(f)\neq0$ or multiple species
rather easily. Solving these equations is important to understand
a range of phenomena in plasma physics \textcolor{black}{(see, e.g.,
\cite{palmroth2018,galeotti2005,manfredi1997,brunetti2000})}. However,
the problem is extremely demanding from a computational point of view,
as it is posed in a high-dimensional (up to six-dimensional) phase
space. Moreover, due the appearance of small scale structures in the
solution (filamentation) a moderate to large number of grid points
is often required to resolve the dynamics of interest. We are interested
in routinely solving four and five dimensional, and eventually even
six-dimensional, problems.

A large body of research exists that discusses solving such kinetic
equations.  Perhaps the most common approach, at least in the physics
literature, is using so-called particle methods (e.g.~particle in
cell schemes; see \cite{verboncoeur2005particle} for a review). There
no direct discretization of velocity space is performed. Instead,
particles are used to represent the kinetic dynamics. The density
function $f$ is then reconstructed from these particles. In the context
of GPU implementations we, in particular, mention \cite{PIConGPU2013,rossi2012towards,crestetto2012resolution}.
However, particle methods have a number of drawbacks. Among them is
that these methods suffer from numerical noise. This often makes it
difficult to discern certain parts of phase space, particularly in
regions where the density is low. An alternative approach is to use
so-called semi-Lagrangian schemes. These methods directly discretize
phase space and consequently do not suffer from numerical noise. For
an overview of the available literature we refer to \cite{palmroth2018}.
\textcolor{black}{We note that semi-Lagrangian schemes are usually
preferred over traditional explicit methods (such as Runge--Kutta
or multi-step schemes); this is due to the relatively stringent CFL
condition, which, for the Vlasov equation, is induced by the largest
velocity in the system (see, e.g., \cite{filbet2001}).}\textcolor{black}{{}
}We will discuss some aspects of semi-Lagrangian schemes that are
pertinent to the present work in the next section. Let us also mention
that recently the use of dimension reduction techniques, such as low-rank
approximations, have been explored \cite{Kormann15,Kormann,El18,EL18_cons,Guo2016,EOP_19}.

We proceed as follows. In section \ref{sec:algorithms-and-code} we
introduce the semi-Lagrangian discontinuous Galerkin algorithm and
discuss our multi-GPU implementation within the SLDG framework. We
then present comparative performance results for a range of GPUs (V100,
Titan V, K80) both in the single GPU, section \ref{sec:single-gpu},
and multi-GPU, section \ref{sec:multiple-gpu}, setting. Single precision
computations are a possibility to extract even more performance for
certain algorithms. We discuss this aspect in section \ref{sec:single-precision}.
Then, the performance of our implementation on consumer level GPUs
is considered in section \ref{sec:consumer-level}. One of the main
features of our code is that, due to the use of C++ templates, we
have only one code base no matter the dimensionality of phase space.
The corresponding performance implications are discussed in section
\ref{sec:1x3v-and-2x3v}. We then provide some examples of numerical
simulations that can be conducted using the SLDG framework and discuss
the verification of the code (section \ref{sec:numerical-simulation}).
Finally, we conclude in section \ref{sec:conclusion}.

\section{Description of the algorithm and code\label{sec:algorithms-and-code}}

The first step in performing simulations on GPUs, or accelerators
more generally, is to select an appropriate algorithm. Not all algorithms
that have been proposed, evaluated, analyzed, and optimized in the
context of performing semi-Lagrangian Vlasov simulations map equally
well to any given hardware. To obtain good performance on GPUs the
following constraints should be kept in mind. First, GPUs require
fine grained parallelism, i.e.~an algorithm that can be decomposed
into a very large number of (almost) independent threads. Moreover,
the algorithm should keep sequential parts to an absolute minimum.
Second, memory access should be as contiguous as possible. Third,
the algorithm should not rely on the presence of large caches. Fourth,
the algorithm should not use excessive amounts of memory, as memory
is a relatively scarce resource on most GPUs. Fifth, the algorithm
should execute with a minimum of branching in performance critical
parts of the code. While most of these criteria, except perhaps for
the presence of large caches, are also important in order to obtain
good performance for a multi-core CPU implementation, GPUs are significantly
less forgiving if some of these constraints are violated. That is,
the corresponding performance hit is usually more severe.\textcolor{black}{{}
The interested reader is referred to barlas2014multicore,CUDABestPractice,drepper2007every
for more details.}

A large number of semi-Lagrangian algorithms have been proposed for
solving the Vlasov equation. Interpolation using cubic splines sonnendrucker1999semi
is a popular approach. Codes such as GYSELA grandgirard2006drift
and SeLaLib selalib are mainly based on this idea. Other algorithms
have been developed as well. For example, those based on the van Leer
scheme \cite{mangeney2002,califano2006,fijalkow1999} or even purely
Eulerian schemes \textcolor{black}{that reduce the computational cost
by} performing mesh refinement \cite{hittinger2013block,jarema2016block,jarema2017block}.
The downside of the schemes based on cubic spline interpolation, and
a range of other algorithms, is that the numerical method can not
be written as simple local update rule. This is due to the fact that
constructing the spline is a global operation with respect to a given
coordinate direction. It also implies that, in order to achieve good
performance, the spline coefficients have to be kept in cache. This
is a minor issue for CPU based systems, where large caches and a relatively
small number of cores are still the norm. However, on GPUs it is simply
not possible that each thread works on a single spline, while keeping
the corresponding coefficients in local memory. Thus, obtaining a
performant implementation in this setting is very difficult. All approaches
to parallelize Eulerian and semi-Lagrangian methods on GPUs have thus
focused on different algorithms \cite{mehrenberger2013vlasov,einkemmer2016,PIConGPU2013,sandroos2013multi}.

Of course, it is very easy to obtain a semi-Lagrangian scheme with
a local update rule. The most basic approach is to use Lagrangian
interpolation instead of splines. However, the main issue with that
approach is that the resulting numerical method is very diffusive,
at least if a stencil of reasonable size is used, and does violate
conservation of mass. Our code employs a semi-Lagrangian approach
based on a discontinuous Galerkin formulation. The algorithm has been
proposed independently by \cite{crouseilles2011,rossmanith2011,qiu2011}.
A more detailed treatise on the particular implementation we employ
can be found in \cite{einkemmer2015}. This semi-Lagrangian discontinuous
Galerkin scheme is mass conservative by construction and has a local
update rule that involves data from at most two adjacent cells \cite{einkemmer2017study}.
The method has been compared, in the four dimensional setting, to
cubic spline interpolation in \cite{einkemmer2019performance}. There
it was found that even on the CPU the method is very competitive (in
certain test examples even significantly faster). The semi-Lagrangian
discontinuous Galerkin approach actually yields a family of methods
with different orders. The method has been studied extensively and
there is even a mathematical convergence analysis available \cite{einkemmer2014}.
Due to all of these reasons, but primarily because of the local update
rule and the competitive performance, we have chosen to base our implementation
on this numerical scheme.

Before presenting the actual results it is warranted to discuss the
framework within which all of this is done. The SLDG code has been
developed over the last few years. The current version can be found
at \url{https://bitbucket.org/leinkemmer/sldg}. For the performance
results in sections \ref{sec:single-gpu}-\ref{sec:1x3v-and-2x3v}
we have used commit \texttt{6f8def0} on the \texttt{master} branch.
The distinct feature of our implementation is that we employ C++ templates
in order to have a single code base for problems in arbitrary dimensions.
That is, we can use the same implementation for 2+2 dimensional (2
dimensions in space and 2 dimension in velocity), 1+3 dimensional
problems, 2+3 dimensional problems, 1+1 dimensional problems, etc.
Templates, in particular, help us to achieve this level of abstraction
while still obtaining good performance. All routines related to advection
or other performance critical parts of the code have three template
parameters: the number of dimensions in space, the number of dimension
in velocity, and the order of the method. This allows the compiler
to generate code that is specific to these situations, which is critical
for obtaining good performance. This is true for both CPU based systems
as well as GPU based systems.

The second major feature of SLDG is that we only have a single implementation
of the semi-Lagrangian discontinuous Galerkin algorithm. This implementation
is used both on the CPU and on the GPU. Only the code that calls the
computational kernel is (slightly) different between the different
hardware platforms. We use OpenMP for the CPU implementation and CUDA
for the GPU implementation. This is accomplished by having a thin
layer over the computational kernel that decides how the different
loop iterations are mapped to the hardware. In CUDA these iterations
are mapped to threads and blocks and on the CPU side they map to traditional
loop iterations, the outermost of which is parallelized using OpenMP.
\textcolor{black}{The OpenMP code uses the first touch principle in
order to exploit the full memory bandwidth on NUMA (Non-uniform memory
access) systems (this is complemented by setting the appropriate OpenMP
environment variables to pin each threads to a fixed core). }We also
apply some optimizations at this level. This is especially true for
the CPU implementation, where we explicitly write out the innermost
loop (i.e.~separate it from the remaining dimension independent loop
construct) to enable the compiler to perform a number of optimizations.
We also implement cache blocking at this level. The GPU implementation
also introduces some complexity as we have to map our problem in a
sensible way to blocks and threads and respect the corresponding constraints
of shared memory. We feel that the present implementation is a good
compromise for obtaining good performance with a maximal level of
abstraction, while still having the ability to relatively easily add
new hardware platforms in the future.

The code also supports running simulation on multiple GPUs (using
both peer-to-peer transfer and data transfer via the host, i.e.~the
CPU) and MPI parallelization. The former is accomplished in terms
of what we call a block. A block, in our nomenclature, is a subset
of the computational domain. There can be multiple blocks for each
MPI process. Each block is then assigned a single GPU and there are
facilities provided so that different blocks on the same shared memory
node can efficiently interchange data (for example, boundary cells),
either directly using peer-to-peer transfer or by copying the data
first to the host and from there to the destination GPU. Obviously,
this facility does \textit{not} use MPI which is very important to
obtain good performance; this is especially true on modern multi-GPU
systems with NVLink or fast PCIe interconnect. The data distribution
to multiple GPUs or MPI processes is done in the velocity directions
only. This is quite common for such codes, see e.g.~\cite{selalib}.

\begin{figure}[H]
\centering{}\includegraphics[width=9cm]{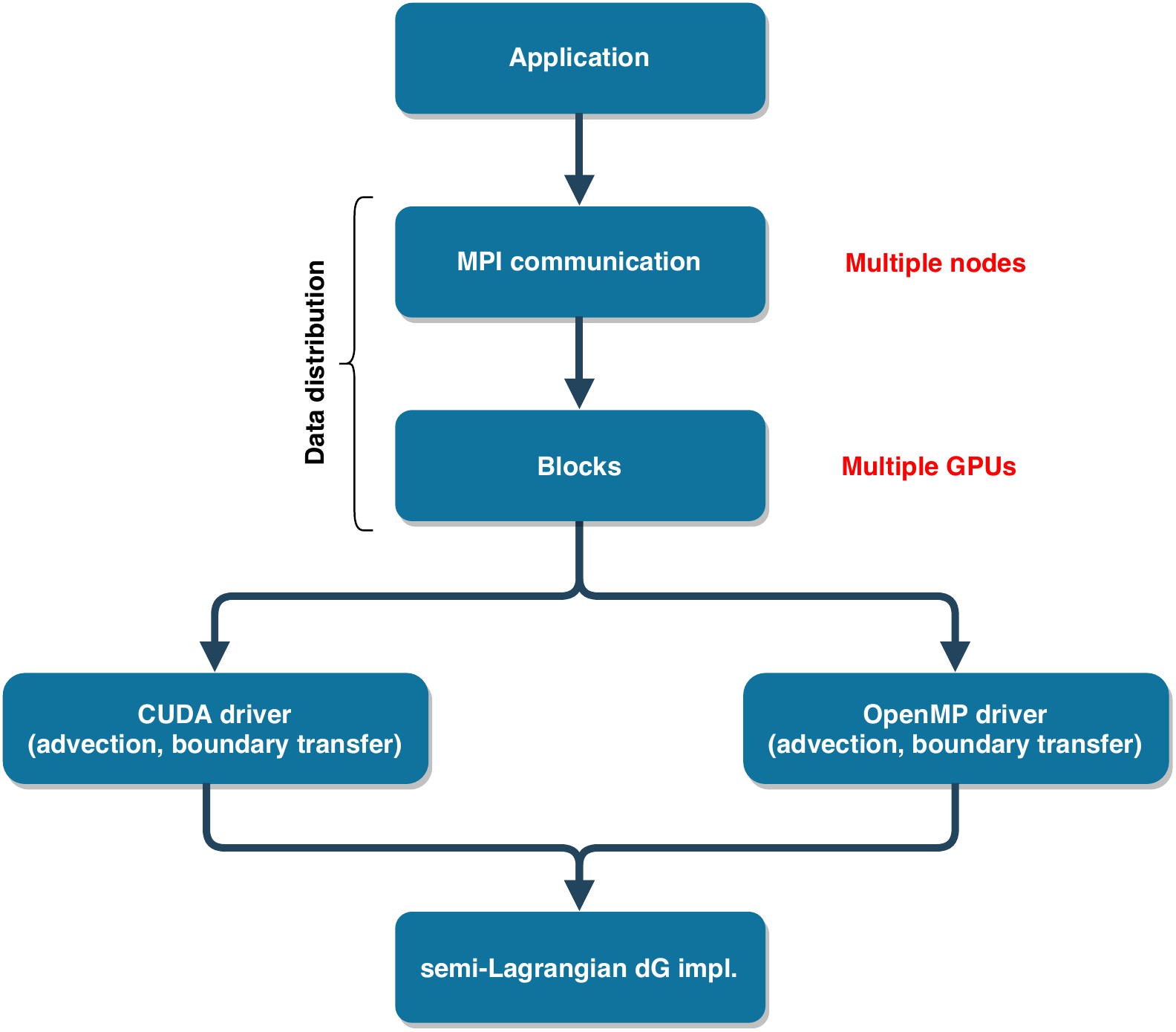}\caption{An overview of the structure of the SLDG code used in this paper.
\label{fig:sldg-structure}}
\end{figure}

An overview of the structure of the code is shown in Figure \ref{fig:sldg-structure}.
Let us also remark that the CPU implementation has been compared to
SeLaLib in \textcolor{black}{\cite{einkemmer2019performance}} and
been found to give better performance. \textcolor{black}{The overall
performance improvement ranges from $25\%$ to a more than an order
of magnitude, depending on the specific problem considered. This comparison
has been conducted on a dual socket Intel Haswell system with a total
of 16 cores. Memory use for SLDG is also reduced by approximately
a factor of three (assuming both methods use the same number of degrees
of freedom).} However, one should be somewhat cautious when interpreting
such comparisons since the goal of SeLaLib is different. SeLaLib is
more generic in the sense that many different semi-Lagrangian schemes
can be implemented, while our code focuses on a specific numerical
method. On other hand, our implementation is independent of the dimension
of the problem, while in the Fortran library SeLaLib a specific program,
supported by the library, is written for each problem with a given
dimension. There are also important differences in the actual implementation
and the design of the two frameworks.

The main computational effort in the present algorithm is performing
the advections and computing the density/physical invariants. For
the former, as mentioned above, a semi-Lagrangian discontinuous Galerkin
scheme is used. The degrees of freedom are given by function values
$u_{ij}^{m}\approx u^{m}(x_{i-1/2}+\xi_{j})$, where $i$ is the cell-index,
which runs from $0$ to $n_{C}$, $x_{i-1/2}$ is the left cell interface,
and $\xi_{j}$ is the $j$th Gauss--Legendre node scaled to the interval
$[0,h]$, where $h$ is the cell size. The indices $j$ run from $0$
to $o-1$, where $o$ is the order of the method. For simplicity we
only consider the one-dimensional case here. More detailed information
can be found in \cite{einkemmer2015}. To perform the $L^{2}$ projection
at the heart of the algorithm only data from at most two adjacent
cells is required. We denote the index of these two cells by $i^{\star}$
and $i^{\star}+1$, respectively. The resulting numerical scheme then
computes $u_{ij}^{m+1}$ from $u_{ij}^{m}$ as follows
\[
u_{ij}^{m+1}=\sum_{l}A_{jl}u_{i^{\star}l}^{m}+\sum_{j}B_{jl}u_{i^{\star}+1;l}^{m},
\]
where $A\in\mathbb{R}^{o\times o}$ and $B\in\mathbb{R}^{o\times o}$.
This computation has to be done for each $i$ and $j$ and for each
slice of the high-dimensional problem. The amount of memory accesses
is equal to two times (one read and one write) the degrees of freedom.
On the other hand, we require at least $2o$ times the degrees of
freedom many arithmetic operations. Nevertheless, this is a memory
bound problem on most modern hardware architectures (including GPUs).
We also mention that assembling the (small) matrices $A$ and $B$
takes some time. However, this usually only requires a small part
of the overall run time. Computing the density and the physical invariants
is also essentially a memory bound problem. However, we note that
to compute the invariants a fair amount of arithmetic operations have
to be performed as well. \textcolor{black}{We are required to compute
the density in every time step as the electric field depends on it.
We also calculate the other physical quantities (such as the electric
energy, kinetic energy, total energy, mass, $L^{2}$ norm, ...) in
every time step. Some of those, e.g. the electric energy, are used
for visualization of the solution, while others, such as the total
energy, serve as a diagnostic for the numerical solution.}

\section{Single GPU performance and comparison with CPU\label{sec:single-gpu}}

In this section we will consider the performance of a single GPU and
compare it to a dual socket CPU system. We consider a NVIDIA V100
GPU (with NVLink interconnect) and a NVIDIA Titan V. Both of these
GPUs are based on the Volta architecture. For comparison we will also
consider the (rather outdated) NVIDIA K80. The K80 is a dual GPU board.
Thus, for the results in this section we will only use one of the
two GPUs that are part of the K80 package. A list of the theoretical
hardware characteristics of these GPUs is given in Table \ref{tab:hardware-specs}.
The main CPU system used for comparison is based on two 16 core Intel
Xeon Gold 6130 CPUs operating at 2.1 GHz. More details can be found
in Table \ref{tab:hardware-specs}.

\begin{table}[H]
\centering{}%
\begin{tabular}{rrrrrr}
\hline 
 &  & \multicolumn{2}{c}{TFlops/s} &  & \tabularnewline
\cline{3-4} \cline{4-4} 
 &  & double & single &  & GB/s\tabularnewline
\hline 
2x Xeon Gold 6130 &  & 2.2 & 4.3 &  & \textcolor{black}{256}\tabularnewline
2x Xeon E5-2630 v3 &  & 0.6 & 1.2 &  & 59\tabularnewline
V100 &  & 7.5 & 15 &  & 900\tabularnewline
Titan V &  & 6.9 & 13.8 &  & 653\tabularnewline
0.5x K80 &  & 1.5 & 4.4 &  & 240\tabularnewline
1x K80 &  & 2.9 & 8.7 &  & 480\tabularnewline
GTX 1080 Ti &  & 0.35 & 11.3 &  & 484\tabularnewline
\textcolor{black}{NVLink (V100)} &  & \textcolor{black}{--} & \textcolor{black}{--} &  & \textcolor{black}{300}\tabularnewline
\textcolor{black}{PCIe 3.0 x16} &  & \textcolor{black}{--} & \textcolor{black}{--} &  & \textcolor{black}{16}\tabularnewline
\hline 
\end{tabular}\caption{The main hardware characteristics, i.e.~peak arithmetic performance
for single and double precision and the theoretically attainable memory
bandwidth, for the dual socket server system and the GPUs that are
used in the numerical simulations are listed. \textcolor{black}{In addition,
we also state the theoretical bandwidth for NVLink (on the V100) and
the theoretical bandwidth for the PCIe configuration found in our
system.} \label{tab:hardware-specs}}
\end{table}

All performance results given in this section will be measured in
terms of \textit{achieved bandwidth. }That is, we count the number
of bytes we have to transfer from (read) and to (write) memory. In
the Strang splitting algorithm used we have to conduct $6$ advections,
which require one read of the input array and one write to the output
array each \textcolor{black}{(this assumes that every duplicated local
memory access incurs no additional cost; i.e., that locally all data
accesses are perfectly cached)}. In addition, we have to compute the
density $\rho$ twice. Once at the middle of the time step to compute
the appropriate electric field to obtain second order and once at
the beginning of the time step to write the electric energy and other
invariant to file. Each of those reductions requires one read of the
input array. Thus, for each time step we have to access the entire
data set in memory $14$ times. The total amount of memory transferred
is $14\cdot N\cdot\text{sizeof(}\text{double})$, where $N$ is the
number of degrees of freedom that are used in the simulation. The
\textit{achieved bandwidth} is then computed as follows
\begin{equation}
\text{achieved bandwidth}=\frac{14\cdot N\cdot\text{sizeof}(\text{double})}{t_{\text{timestep}}},\label{eq:achieved-bandwidth}
\end{equation}
where $t_{\text{timestep}}$ is the average wall-time required to
compute one time step. In the actual algorithm a number of additional
tasks have to be done. For example, computing the electric field from
the density, copying the electric field to the GPU, assembling boundary
conditions and copying them to the GPU, etc. However, since those
operations are usually cheap compared to operations that involve the
entire data set, we neglect them for the present consideration.

In contrast to the also very common practice of stating floating point
operations per second, using the achieved bandwidth to measure performance,
and thus as our figure of merit, has two main advantages. First, the
numerical method is memory bound (see the discussion above and the
references given). Thus, stating the performance in terms of bandwidth
makes it easy to compare the performance to what is theoretically
possible on the underlying hardware. Second, the measure is independent
of the order of the numerical method. That is, no matter if we use
a second or fourth order scheme the amount of data we have to read/write
only depends on the total number of degrees of freedom. In addition,
the way we have defined achieved bandwidth here does not depend on
whether we use a CPU, a single GPU, or multiple GPUs. The achieved
bandwidth is proportional to the inverse of the wall time (this is
possibly since we do not take, for example, copying boundary data
between GPUs into account; this is all overhead that reduces the performance
and is not considered part of the algorithm). This makes comparisons
in terms of what practitioners care about, i.e.~wall time, straightforward.
\textcolor{black}{It should, however, be noted that the achieved memory
bandwidth can not be directly used to compare different numerical
methods; in particular, this applies to methods with different order.
In this case any advantage in achieved bandwidth would have to be
contrasted to a potential decrease in accuracy. Weather a higher method
or a lower order method is more effective also depends strongly on
the specific problem, weather we are interested in long time behavior
or not, as well as many other factors. For the semi-Lagrangian discontinuous
Galerkin scheme some of those factors have been considered in \cite{einkemmer2019performance,einkemmer2017study}.
We refer the interested reader to these papers for more details. However,
for our present goal, i.e.~evaluating the efficiency of the implementation
using the achieved bandwidth as the figure of merit is the appropriate
approach.}

\begin{figure}[H]
\begin{centering}
\includegraphics[width=7cm]{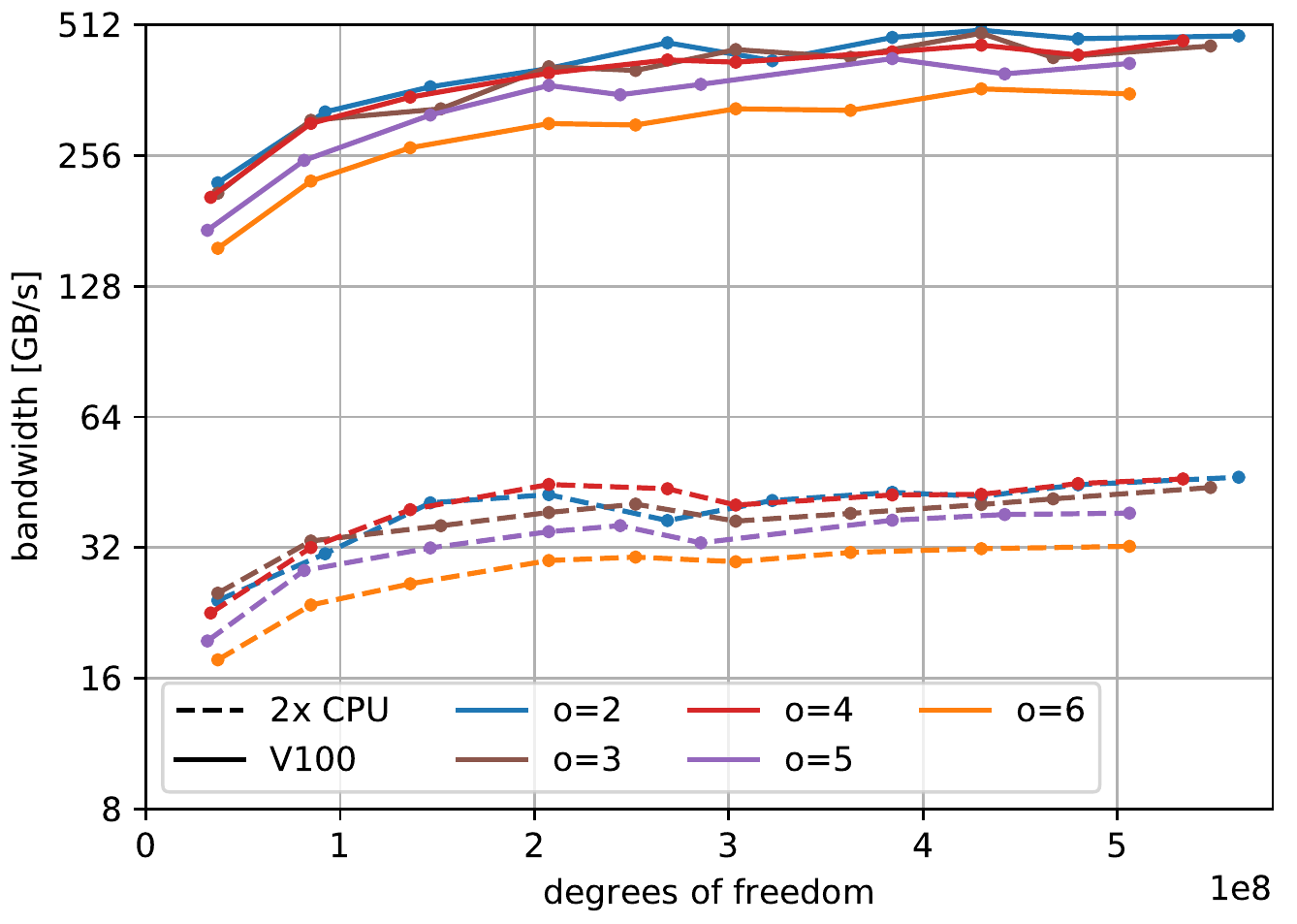}\hspace{1cm}\includegraphics[width=7cm]{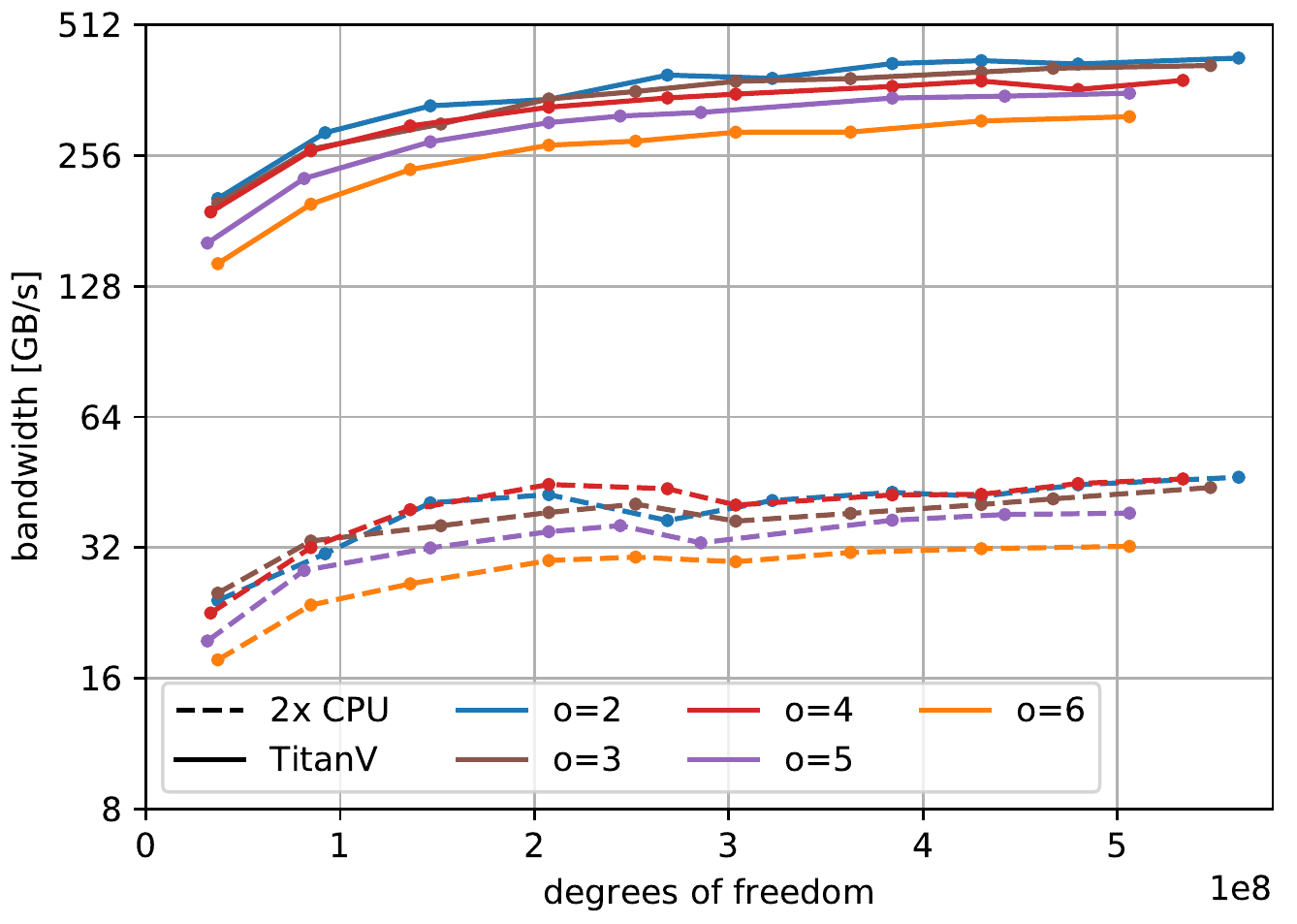}
\par\end{centering}
\caption{Single GPU and dual socket CPU performance for solving the 2+2 dimensional
Vlasov--Poisson equation. The achieved bandwidth for a V100 (left)
and Titan V (right) as a function of the number of degrees of freedom
is shown. The simulation are performed using double precision floating
point numbers. \label{fig:single-gpu-tesla}}
\end{figure}

We consider the integration of a four dimensional Vlasov--Poisson
equations (two dimensions in physical space and two dimensions in
velocity) on a single GPU/dual socket CPU system. The corresponding
results are shown in Figure \ref{fig:single-gpu-tesla}. The performance
is investigated as a function of the number of degrees of freedom
(on the $x$-axis) and the order $o$ of the numerical method. On
the CPU the performance is at most $46$ GB/s. For the V100 GPU we
observe between 355 (order 6) and 483 (order 2) GB/s. There is a significant
penalty in terms of \textcolor{black}{achieved bandwidth }when using
higher order methods. However, we should note that the 4th order method,
which seems to be a good choice for many problems in practice, still
achieves 470 GB/s on the V100 (and 381 GB/s on the Titan V).\textcolor{black}{{}
A simple streaming benchmark (using a call to }\texttt{\textcolor{black}{cudaMemcpy)}}\textcolor{black}{{}
on the V100 gives approximately 800 GB/s (significantly less than
the 900 GB/s listed by the vendor). This is the maximum a memory bound
problem can achieve. Our implementation obtains approximately 60\%
of that value. Although, research on stencil codes has demonstrated
results that are closer to this limit \cite{pershin2019,rawat2018,rodrigues2019gpu},
given that our numerical algorithm needs significantly more floating
point operations, operates in a higher dimensional setting (almost
all stencil implementations consider only problems in up to three
dimensions), is implemented independently of the dimension of the
problem (which is not true for most stencil codes which either have
a fixed dimension \cite{pershin2019} or use code generation \cite{rawat2018,rodrigues2019gpu}),
and the fact that we include overhead specific to the Vlasov equation
(in particular, the Poisson solver and the communication of boundary
data) in computing the achieved bandwidth, means that we achieve excellent
performance.}

Overall, there is approximately a factor of $10$ speed up for the
V100 compared to our dual socket CPU node. \textcolor{black}{We note
that the percentage of the theoretical peak performance achieved for
the CPU implementation is somewhat less compared to the GPU implementation.
In our opinion this is ultimately due to the simpler architecture
of the GPU, which gives us the ability to more directly control data
transfer (e.g.~via shared memory). Generally, we also found the performance
on the GPU to be more predictable. On the CPU side even some relatively
minor changes could have a significant impact on performance. Moreover,
although the problem is memory bound, a significant amount of arithmetic
operations still has to be performed. In our experience GPUs are better
able to hide the associated latency. }Let us also note that the performance
characteristics for the Titan V are rather similar to the V100; the
code achieves between 315 (order $6$) and 429 GB/s (order $2$).
The performance difference between the two platforms is between 10
and 25\%, depending on the order of the method.

\begin{figure}[H]
\begin{centering}
\includegraphics[width=7cm]{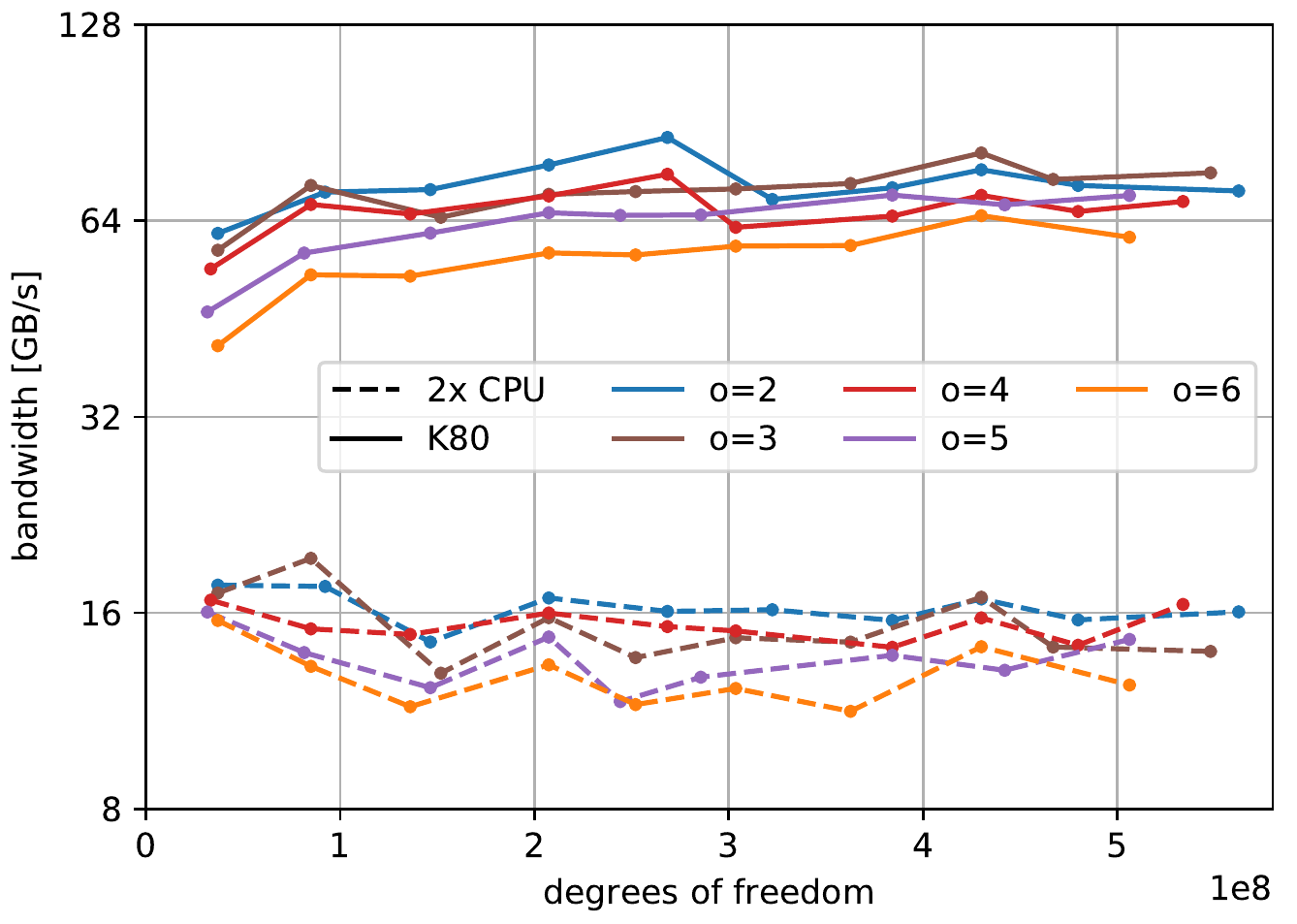}
\par\end{centering}
\caption{Single GPU and dual socket CPU performance for solving the 2+2 dimensional
Vlasov--Poisson equation. The achieved bandwidth for one GPU in the
K80 package is shown as a function of the number of degrees of freedom.
The simulations are performed using double precision floating point
numbers.\label{fig:single-gpu-k80}}
\end{figure}

For historical reasons and to provide further context for the present
comparison, we will also consider a single GPU of the K80 package
(Kepler architecture). Obviously, this is a different system and we
thus use the dual socket Intel Xeon CPU E5-2630 v3 as the basis for
the comparison. The results are shown in Figure \ref{fig:single-gpu-k80}.
Here we achieve approximately 16 GB/s on the dual socket CPU system
and from 60 to 76 GB/s on the GPU. Thus, we observe a speed up on
the GPU of approximately a factor of $4$. The difference in performance
between the V100 and the K80 is approximately a factor $6$ (the release
dates of these two GPUs are approximately three years apart).

\section{Multiple GPU performance\label{sec:multiple-gpu}}

\textcolor{black}{In this section we will consider simulations that
run on multiple GPUs. In this context the entire domain is equally
distributed along the two directions in velocity. Each GPU that takes
part in the computation is thus responsible for a part of the domain.
The algorithm requires an additional communication step that transfers
the required boundary data between the different GPUs. If we denote
the degrees of freedom stored on a single GPU in the $x_{1}$, $x_{2}$,
$v_{1}$, and $v_{2}$ directions by $n_{x_{1}}$, $n_{x_{2}}$, $n_{v_{1}}$,
and $n_{v_{2}}$, respectively, the total number of degrees of freedom
per GPU is $N_{GPU}=n_{x_{1}}n_{x_{_{2}}}n_{v_{1}}n_{v_{2}}$. This
is also the amount of degrees of freedom we have to transfer to and
from memory in every advection step. On the other hand, the amount
of boundary data we have to transfer is significantly smaller. For
example, for the advection in the $x$-direction we have to transfer
$Con_{x_{2}}n_{v_{1}}n_{v_{2}}$ degrees of freedom from each GPU
to each other GPU, where $C$ is the CFL number and $o$ is the order
of the method used, . Thus, the data that has to be transferred is
smaller than the number of memory accesses by a factor of $Co/n_{x_{1}}$.
However, we also have to realize that the bandwidth available is significantly
reduced. In particular, this is true for the systems with PCIe, as
is illustrated in Table \ref{tab:hardware-specs}.}

In this section we consider performance in the multi-GPU, but single
node setting. This is interesting as for both large supercomputers
and desktop workstations it is very common to have multiple GPUs available
on each node. These GPUs are then connected either via NVLink (this
is an option for the V100) or using PCIe. Thus, one main goal of this
section is to investigate the performance implications of NVLink for
our code. The main utility of having a faster interconnect, in the
present context, is to speed up the transfer of the boundary data.Walker

\begin{figure}[H]
\begin{centering}
\includegraphics[width=7cm]{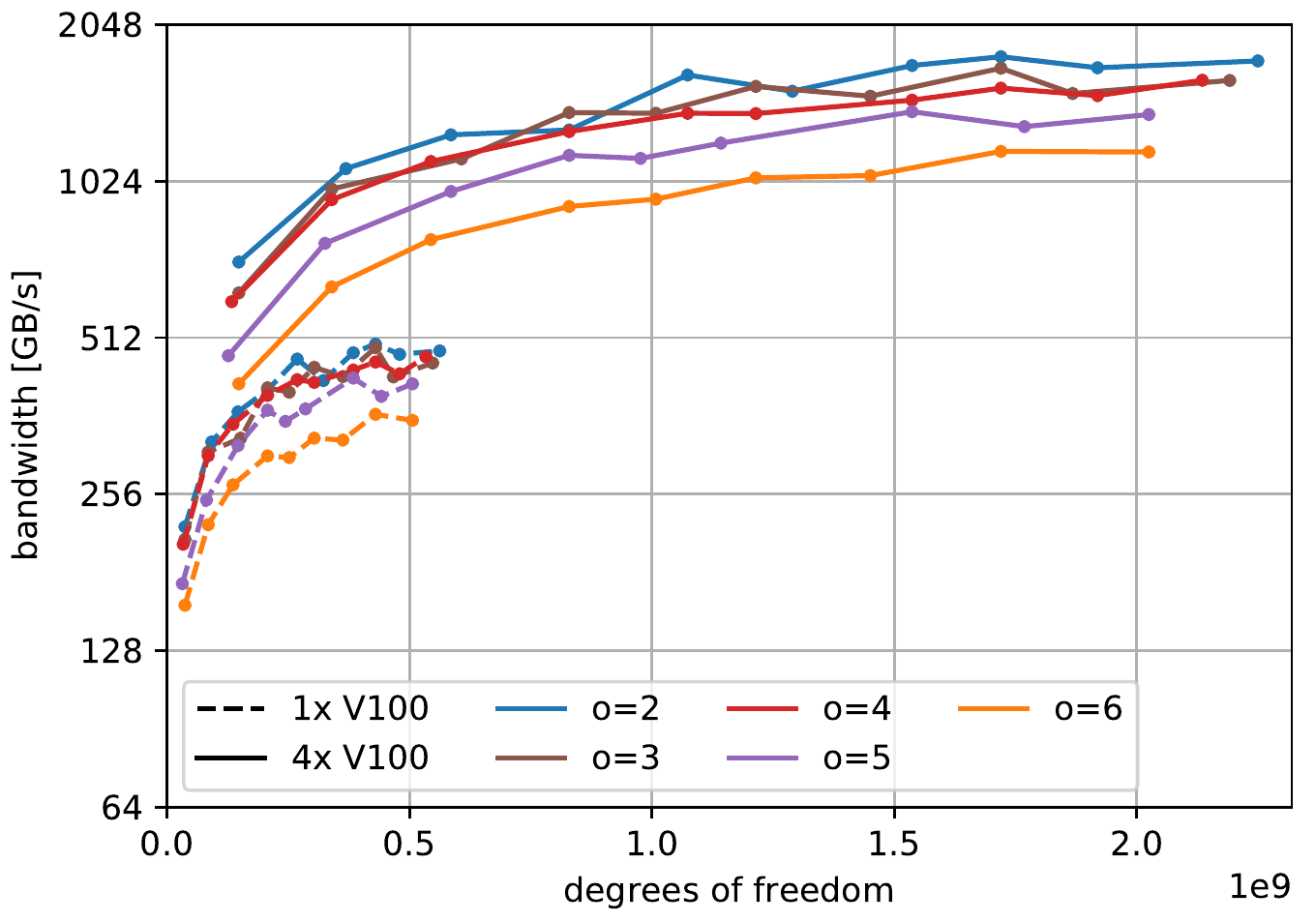}\hspace{1cm}\includegraphics[width=7cm]{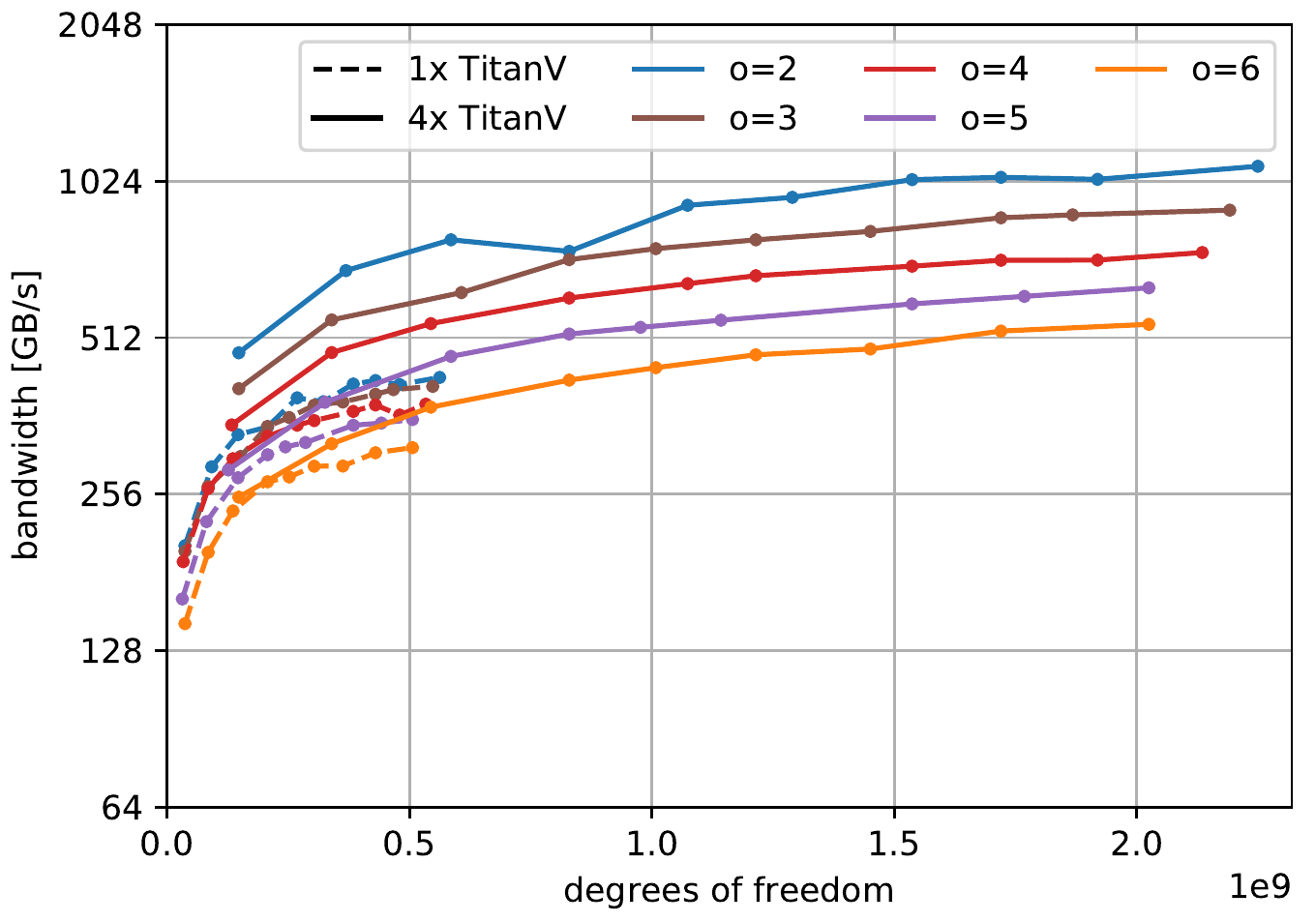}
\par\end{centering}
\caption{Performance of four GPUs on a single node for solving the 2+2 dimensional
Vlasov--Poisson equation. The achieved bandwidth for four V100 (left)
and four Titan V (right) as a function of the number of degrees of
freedom is shown. The simulation are performed using double precision
floating point numbers. For comparison the results for a single GPU
of the same type are shown as well.\label{fig:multiple-gpu-tesla}}
\end{figure}

Our setup consists of four GPUs of each type on the same node. The
V100 GPUs have an NVLink interconnect, while the Titan V are connected
via PCIe. The results for the V100 and Titan V are shown in Figure
\ref{fig:multiple-gpu-tesla}. The performance observed ranges from
1165 GB/s to 1745 GB/s for the V100 and from 543 to 1094 GB/s for
the Titan V. The fourth order method achieves 1602 GB/s on the four
V100 and 747 GB/s on the four Titan V. Thus, we conclude that the
availability of NVLink gives a significant boost in performance, approximately
a factor of two. The speed up compared to a single GPU is also rather
good for the V100 configuration, approximately a factor of $3.5$.
For the Titan V on the other hand, we only observe slightly more than
a factor of two. Thus, there is a significant discrepancy in performance
between the V100 and Titan V configurations.; approximately a factor
of two for the fourth order method.

\begin{figure}[H]
\begin{centering}
\includegraphics[width=8cm]{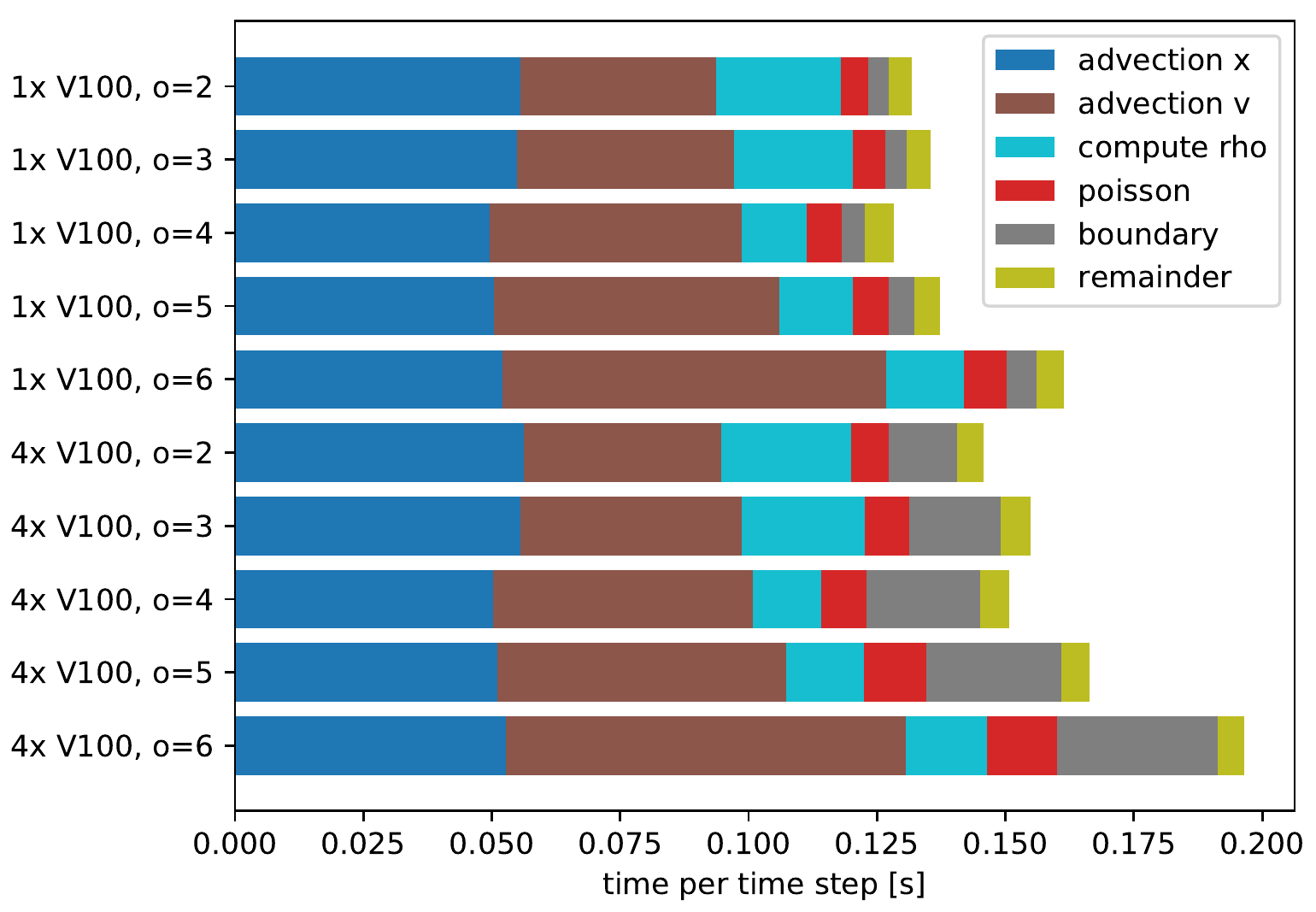}\hspace{0.5cm}\includegraphics[width=8cm]{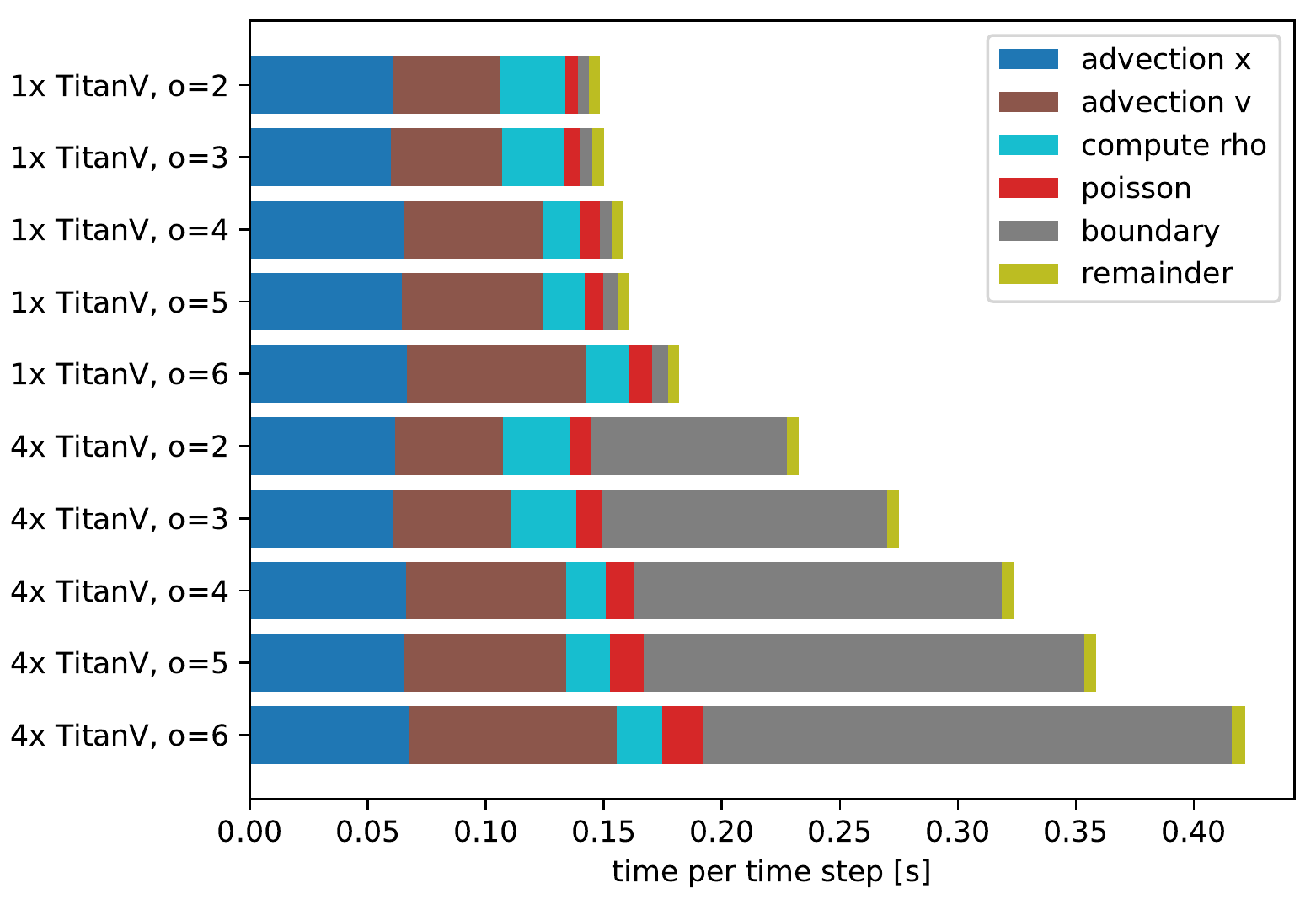}
\par\end{centering}
\caption{The wall time for different parts of the Vlasov solver are shown.
Results are given for single GPU and multi-GPU configurations (V100
on the left and Titan V on the right) and as a function of the order
$o$ of the numerical scheme.\textcolor{black}{{} The degrees of freedom
on each GPU is held constant (strong scaling). That is, the overall
problem size is proportional to the number of GPUs employed (approximately
$0.5\cdot10^{9}$ degrees of freedom for the single GPU setup and
$2\cdot10^{9}$ degrees of freedom for the four GPU setup). Thus,
ideally the time per time step going from a single to four GPUs would
remain constant.}\textcolor{black}{{} }The legend is explained in more
detail in the main text. \label{fig:barchart-o-mult}}
\end{figure}

This is a good time to pause and investigate the performance characteristics
of the code in more detail. To do that we will consider Figure \ref{fig:barchart-o-mult}.
There the wall time for different parts of the program is shown as
a bar chart. We are particularly interested in the disparity in performance
for schemes of different order. We will discuss this both for the
single and the multi-GPU configuration. Figure \ref{fig:barchart-o-mult}
divides the execution of the program into the following parts \textit{advection
x} (computing the translation in the direction of space variables),
\textit{advection v} (computing the translation in the direction of
velocity variables), \textit{compute rho} (computing the density,
required for computing the electric field, and some physical invariants),
\textit{poisson} (computing the electric field by solving a Poisson
equation), \textit{boundary} (collecting the boundary data and all
communication that is necessary), and \textit{remainder} (everything
that has not been accounted for in the listed categories). Before
proceeding, let us note that computing the advection in the velocity
directions is actually more expensive than computing the advection
in the space directions. The bars denote the entire run time and,
due to the the Strang splitting, the advections in the space directions
are called twice as often as the corresponding advections in the velocity
direction.

We clearly see that although there is some additional cost to compute
the advections for higher order schemes, this is only part of the
performance difference. In the multi-GPU setting the time it takes
to transfer boundary data from one GPU to its neighboring GPU is also
very important. This scales unfavorably with the order of the method
as the algorithm has to send a fixed number of cells. For higher order
methods each cell contains more degrees of freedom and thus a larger
amount of data needs to be transferred. This is, of course, particularly
an issue for the slower interconnect on the Titan V and we indeed
see from \ref{fig:barchart-o-mult} that for four Titan V and $o=6$
approximately half of the run time of the algorithm is spend in transferring
boundary data. \textcolor{black}{However, we should note that higher
order methods are also more accurate (for the same number of degrees
of freedom), which at least to some extent compensates for this overhead.}

All configurations so far have used peer-to-peer transfer to directly
copy boundary data from one GPU to another. We will now investigate
how much performance improvement this approach yields compared to
doing the boundary data transfer via the host. There is not really
a point in doing this for the present application, as peer-to-peer
transfer is already implemented and is available on both the V100
and Titan V. Nevertheless, using host transfer is somewhat simpler
and in some settings it could be questioned if the additional complexity
in the code is justified. Moreover, it is interesting to see how much
gain in performance we actually obtain by such an implementation.
The results are shown in Figure \ref{fig:multiple-gpu-tesla}. The
performance difference in our code is quite significant. For example,
for the fourth order method the performance of the V100 drops from
approximately 1602 GB/s (peer-to-peer) to approximately 726 GB/s (host
transfer). For the Titan V the difference is somewhat smaller, as
one might suspect. But even in this case a significant fraction of
performance is lost.

\begin{figure}[H]
\begin{centering}
\includegraphics[width=7cm]{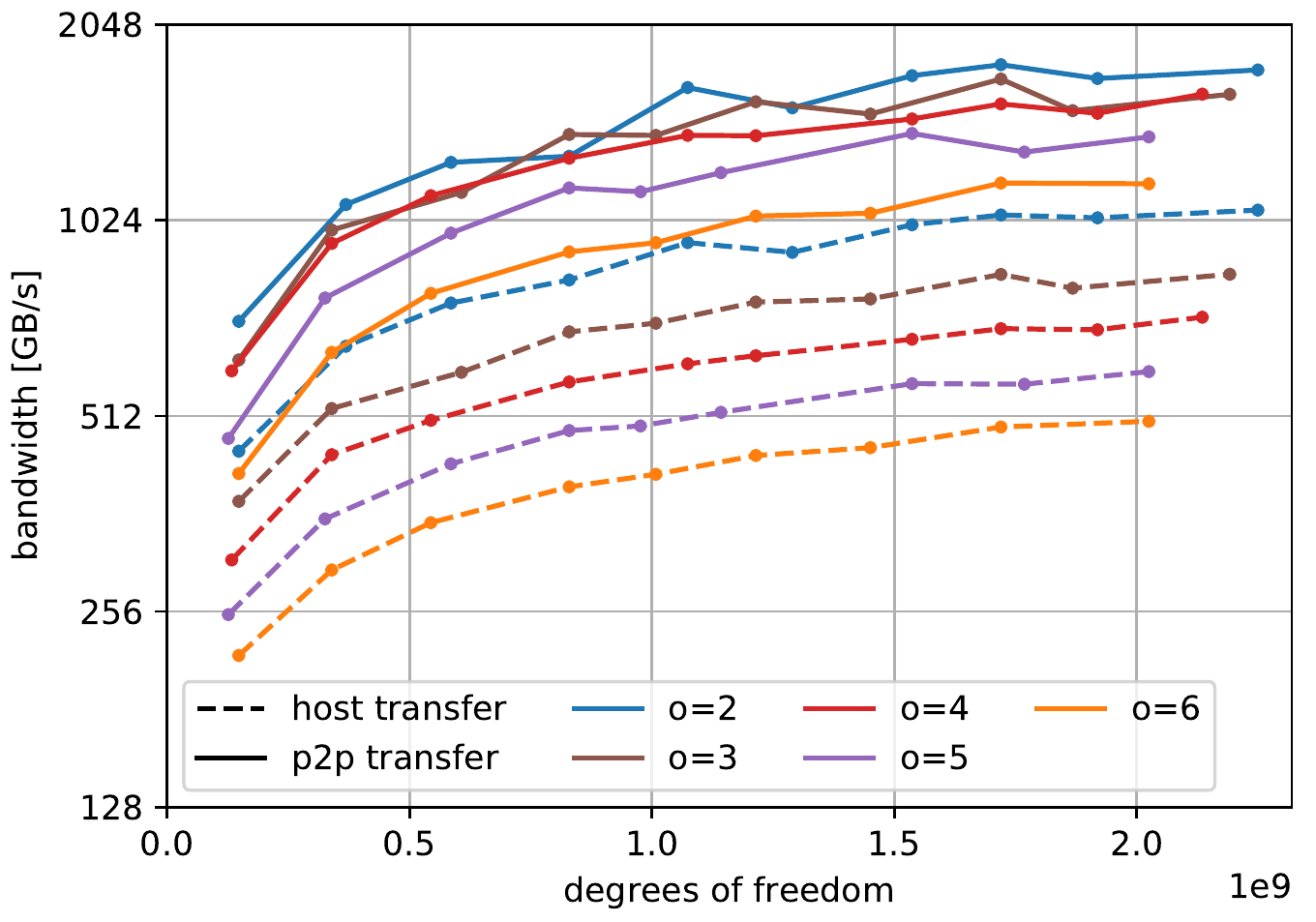}\hspace{1cm}\includegraphics[width=7cm]{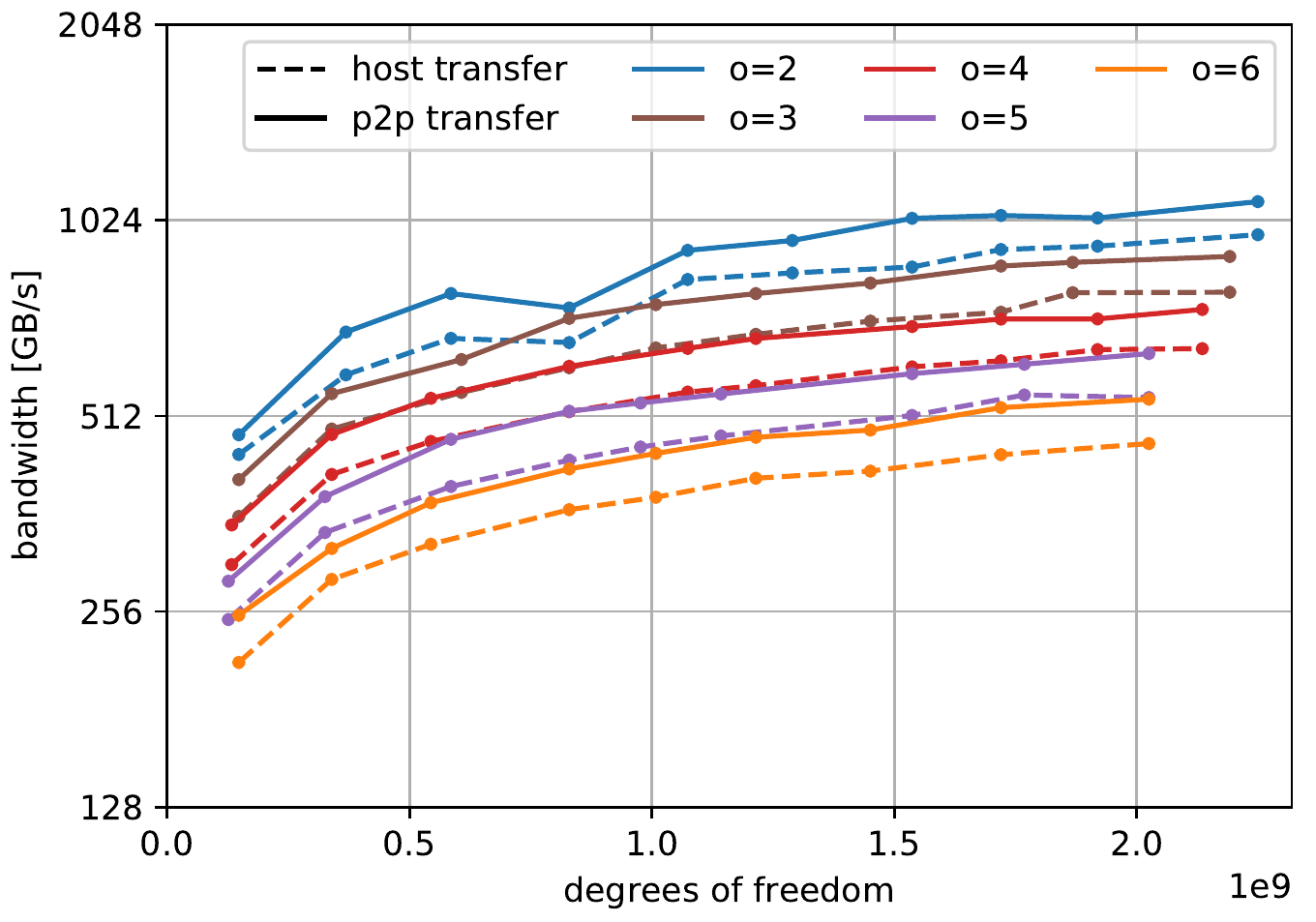}
\par\end{centering}
\caption{Comparison of peer-to-peer data transfer and data transfer via the
host for the 2+2 dimensional Vlasov--Poisson equation. The achieved
bandwidth for four V100 (left) and four Titan V (right) as a function
of the number of degrees of freedom is shown. The simulation are performed
using double precision floating point numbers.\label{fig:p2p-gpu-tesla}}
\end{figure}

\begin{figure}[H]
\begin{centering}
\includegraphics[width=7cm]{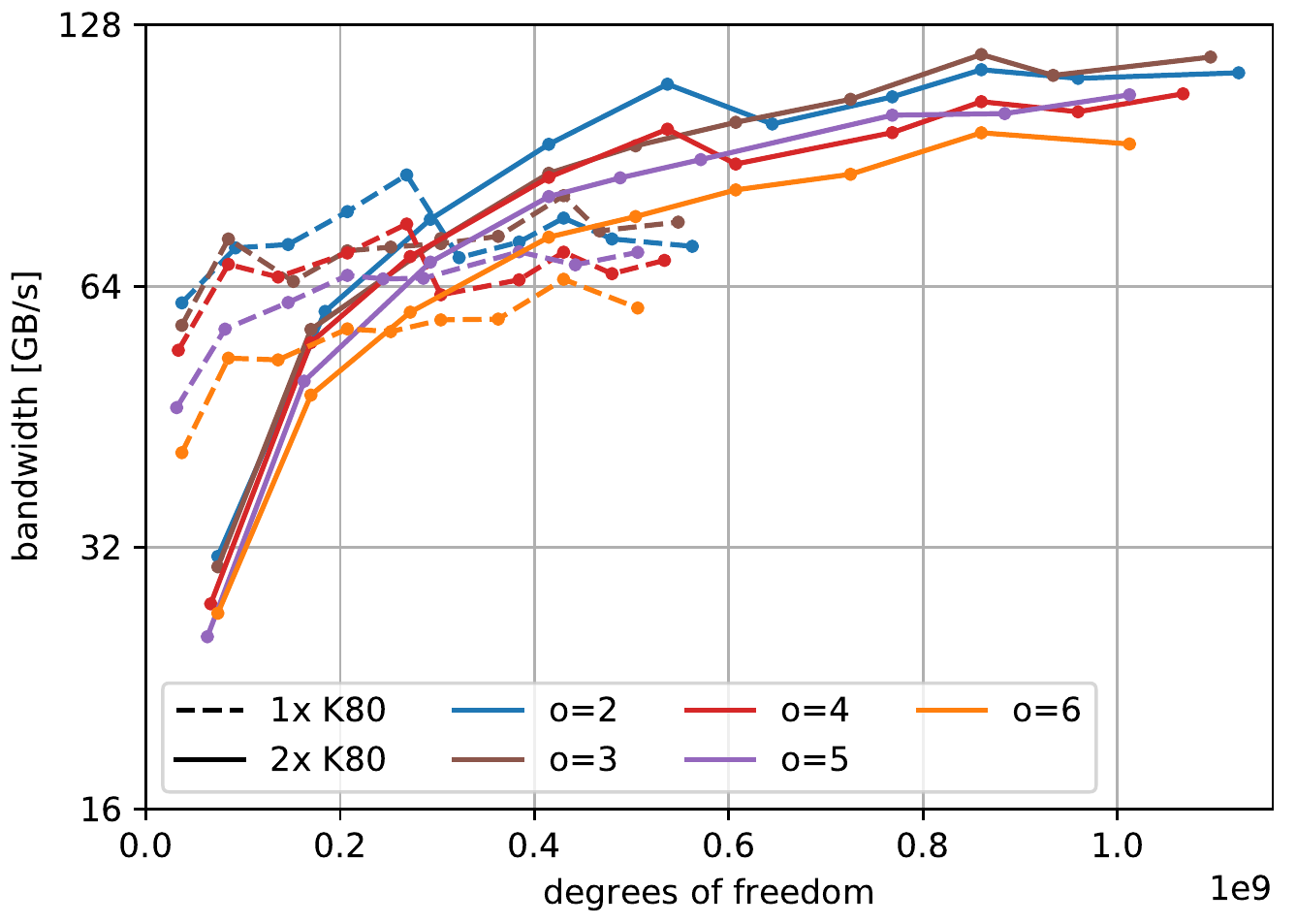}
\par\end{centering}
\caption{Performance of the two GPUs in the K80 package for solving the 2+2
dimensional Vlasov--Poisson equation. The achieved bandwidth as a
function of the number of degrees of freedom is shown. The simulation
are performed using double precision floating point numbers. For comparison
the results for a single GPU are shown as well.\label{fig:multiple-gpu-K80}}
\end{figure}

As before, we also consider the K80. Here we only scale up to the
2 GPUs found in that package. The corresponding results are shown
in Figure \ref{fig:multiple-gpu-K80}. We observe approximately a
factor of $1.5$-$1.7$ improvement in performance by going from a
single GPU to two GPUs. Let us note that even though the two GPUs
in the K80 package are connected using PCIe, this is completely done
within the package provided by the vendor.

\section{Single precision performance\label{sec:single-precision}}

\begin{figure}[H]
\begin{centering}
\includegraphics[width=7cm]{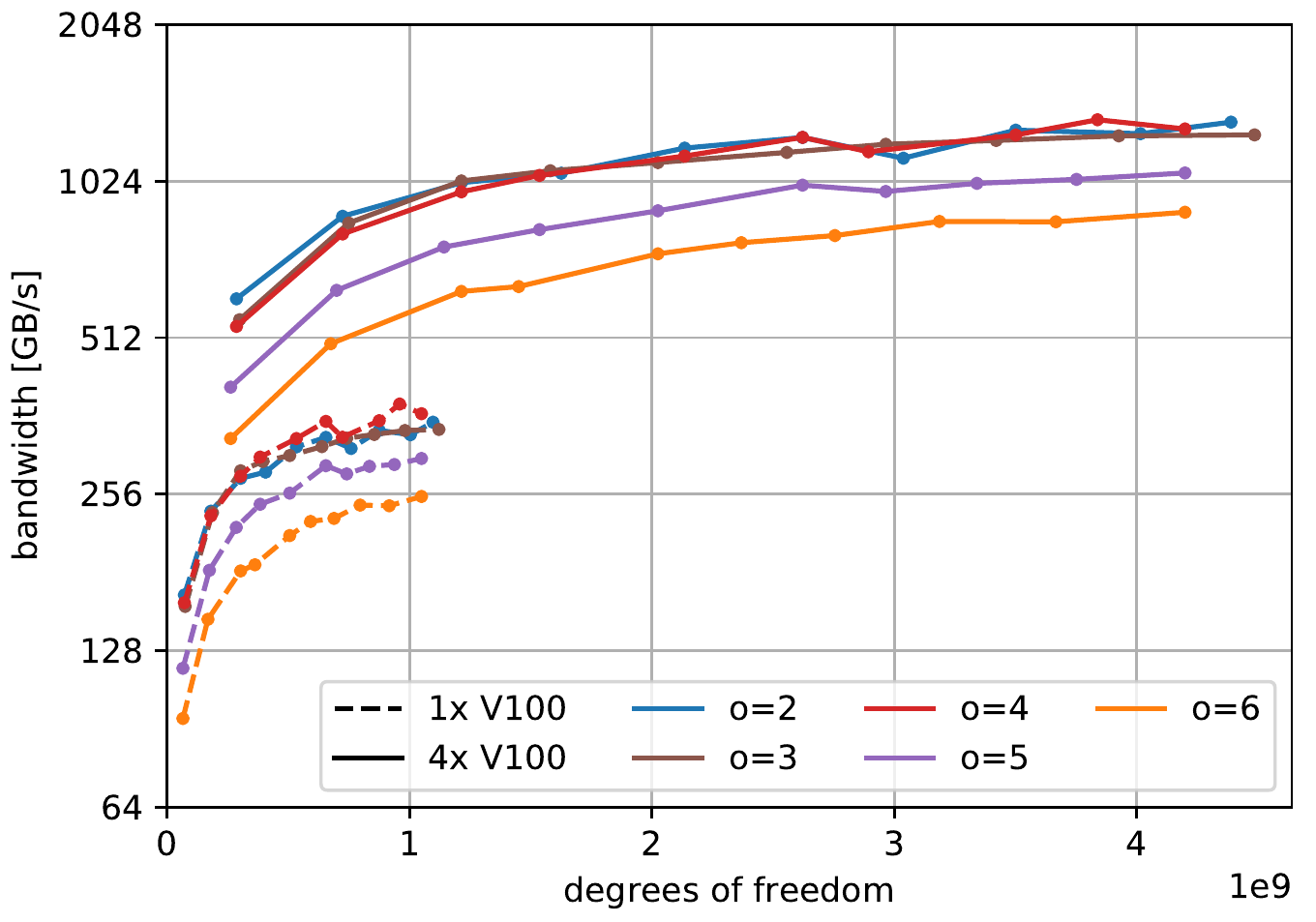}\hspace{1cm}\includegraphics[width=7cm]{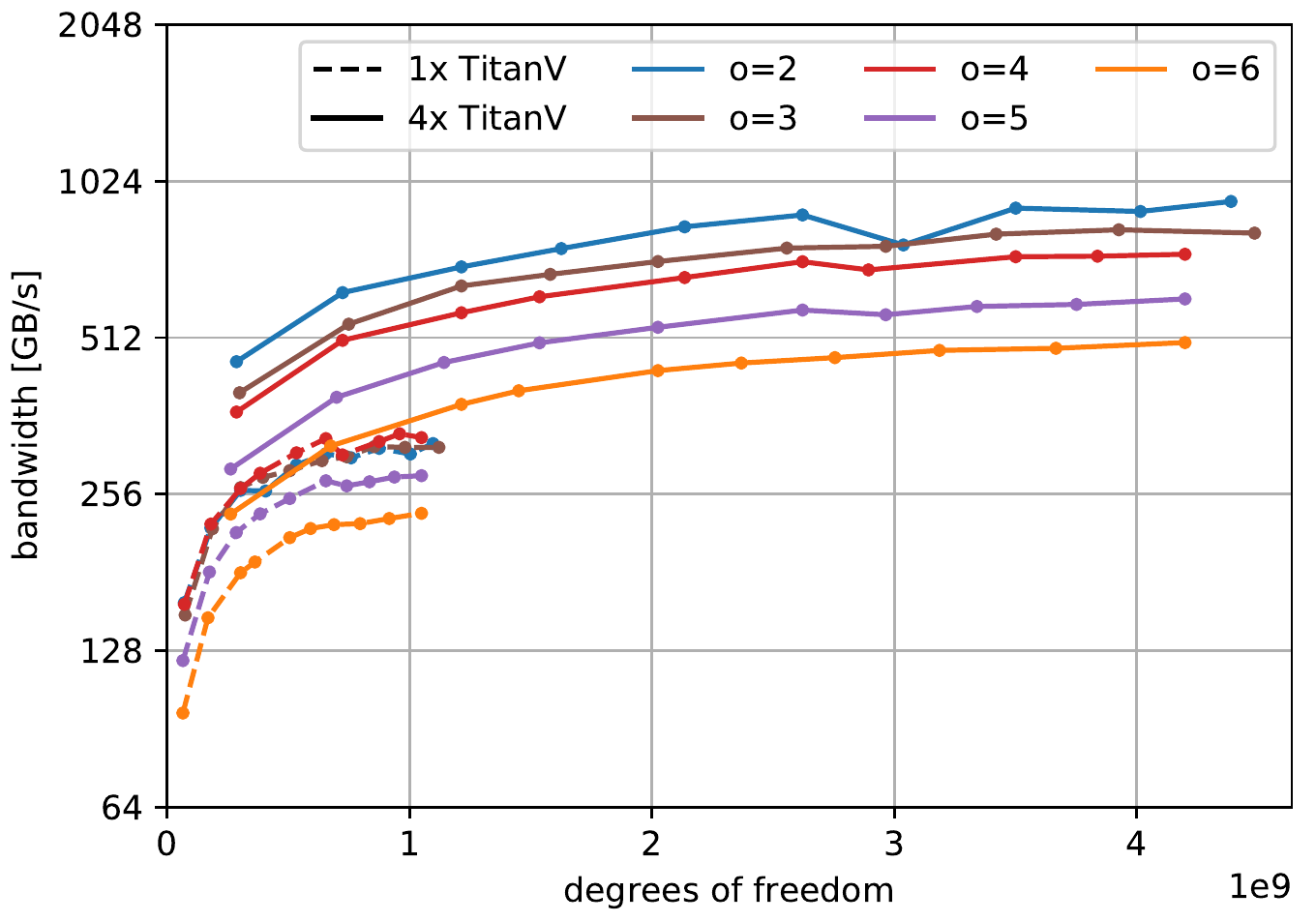}
\par\end{centering}
\caption{Performance of a single and four GPUs on one node for solving the
2+2 dimensional Vlasov--Poisson equation. The achieved bandwidth
for the V100 (left) and the Titan V (right) GPUs as a function of
the number of degrees of freedom is shown. The simulation are performed
using single precision floating point numbers. For comparison the
results for a single GPU of the same type are shown as well.\label{fig:singleprec-multiple-gpu}}
\end{figure}

\begin{figure}[H]
\begin{centering}
\includegraphics[width=7cm]{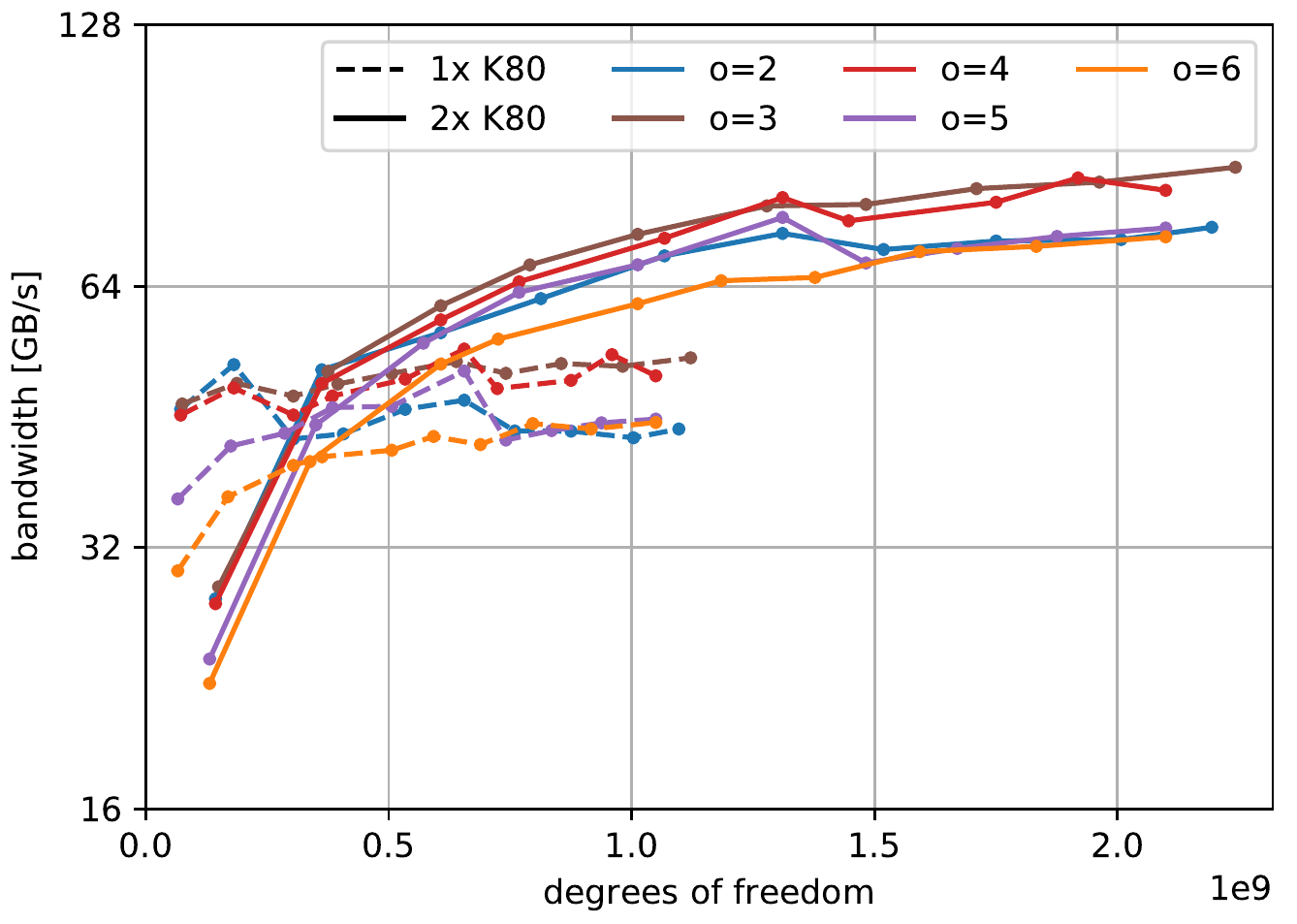}
\par\end{centering}
\caption{Performance of the two GPUs in the K80 package for solving the 2+2
dimensional Vlasov--Poisson equation. The achieved bandwidth as a
function of the number of degrees of freedom is shown. The simulation
are performed using single precision floating point numbers. For comparison
the results for a single GPU are shown as well.\label{fig:singleprec-multiple-K80}}
\end{figure}

Since many scientific applications (stencil codes, sparse matrix operations,
the semi-Lagrangian discontinuous Galerkin algorithm considered here,
etc.) are now memory bound on virtually all modern hardware architectures,
reducing the amount of memory that is required to run a simulation
is beneficial in two ways. First, it increases performance by reducing
the amount of data that have to be read/written to/from memory. Second,
reducing the overall amount of memory can be quite valuable on more
memory constraint platforms. such as GPUs. This enables us to run
a larger simulation on each GPU. Moreover, consumer level GPUs, which
are significantly less expensive, have much better single precision
than double precision performance (we will discuss this more in section
\ref{sec:consumer-level}). Thus, single and mixed precision computation
have become an important topic of research in the last decade. In
fact, for the Vlasov equation a mixed precision algorithm has been
proposed in \cite{einkemmer2016}. The present code supports simulation
using both double and single precision. In particular, care has been
taken that no unnecessary round-off errors are introduced. This is
less an issue for the transport part of the algorithm. Computing a
number of physical quantities, on the other hand, is more problematic
as those requires a reduction over the entire data set. In our implementation
double precision accumulators are used on the CPU. This has only a
negligible impact on performance. On the GPU a natural way to implement
reduction is in a hierarchical fashion (reduction within a single
block first, then reduction within a grid, then reduction of the results
on the CPU). This algorithm is less problematic with respect to round-off
errors. The software package has a number of unit tests that check
if the error of the implementation is reasonable.
\begin{flushleft}
We now consider the same configuration as in section \ref{sec:single-gpu},
except that all data are stored as single precision floating point
numbers. Equation \ref{sec:single-gpu} is modified in the obvious
way to take this into account. The corresponding results are shown
in Figure \ref{fig:singleprec-multiple-gpu} (dashed lines). We observe
that the achieved bandwidth for single precision computations on a
single V100 is somewhat reduced; approximately 15\% for the fourth
order method. For the Titan V the situation is similar. Let us emphasize
that even though the efficiency (as measured by the achieved bandwidth)
is slightly lower, the \textit{real world performance }, i.e. how
many degrees of freedom can be processed in one second, is still significantly
higher. For the fourth order method we observe a reduction in memory
by a factor of two and, for the same number of degrees of freedom,
a reduction in run time by a factor of approximately $1.7$ for the
V100 and $1.8$ for the Titan V. For comparison Figure \ref{fig:singleprec-multiple-K80}
(dashed lines) shows the single precision performance of a single
GPU in the K80 package.
\par\end{flushleft}

We now report single precision results on multiple GPUs, i.e. the
configuration in section \ref{sec:multiple-gpu}. The results for
V100 and Titan V are given in Figure \ref{fig:singleprec-multiple-gpu}
and the corresponding results for the K80 are given in Figure \ref{fig:singleprec-multiple-K80}.
For the V100 the results are very similar to the single GPU case.
That is, we observe a small reduction in efficiency but still an overall
gain in performance by going from double to single precision. For
the Titan V the difference in efficiency between single and double
precision is very small.

\section{Performance of consumer level GPUs\label{sec:consumer-level}}

Consumer GPUs, i.e.~those GPUs that are are primarily marketed towards
gaming applications, have one decisive advantage; namely, their low
price. Their main drawback for scientific computation is the lack
of good performance for double precision computations. Double precision
computations have long been considered the gold standard for doing
scientific computing. However, due to the appearance of hardware architectures
where single precision computations are advantageous, significant
research into single and mixed precision algorithms has been conducted.
Consumer GPUs also do not feature error correcting (ECC) memory nor
do they offer NVLink support. Other than that, the specification of
those GPUs are often almost on par with the Tesla (and Titan V) cards,
the latter of which are more marketed towards the scientific computing
community. Because of this, many fields within scientific computing
that do not require strong double precision performance or very fast
interconnects, such as machine learning or molecular dynamics, have
embraced consumer level GPUs.

\begin{figure}[H]
\begin{centering}
\includegraphics[width=7cm]{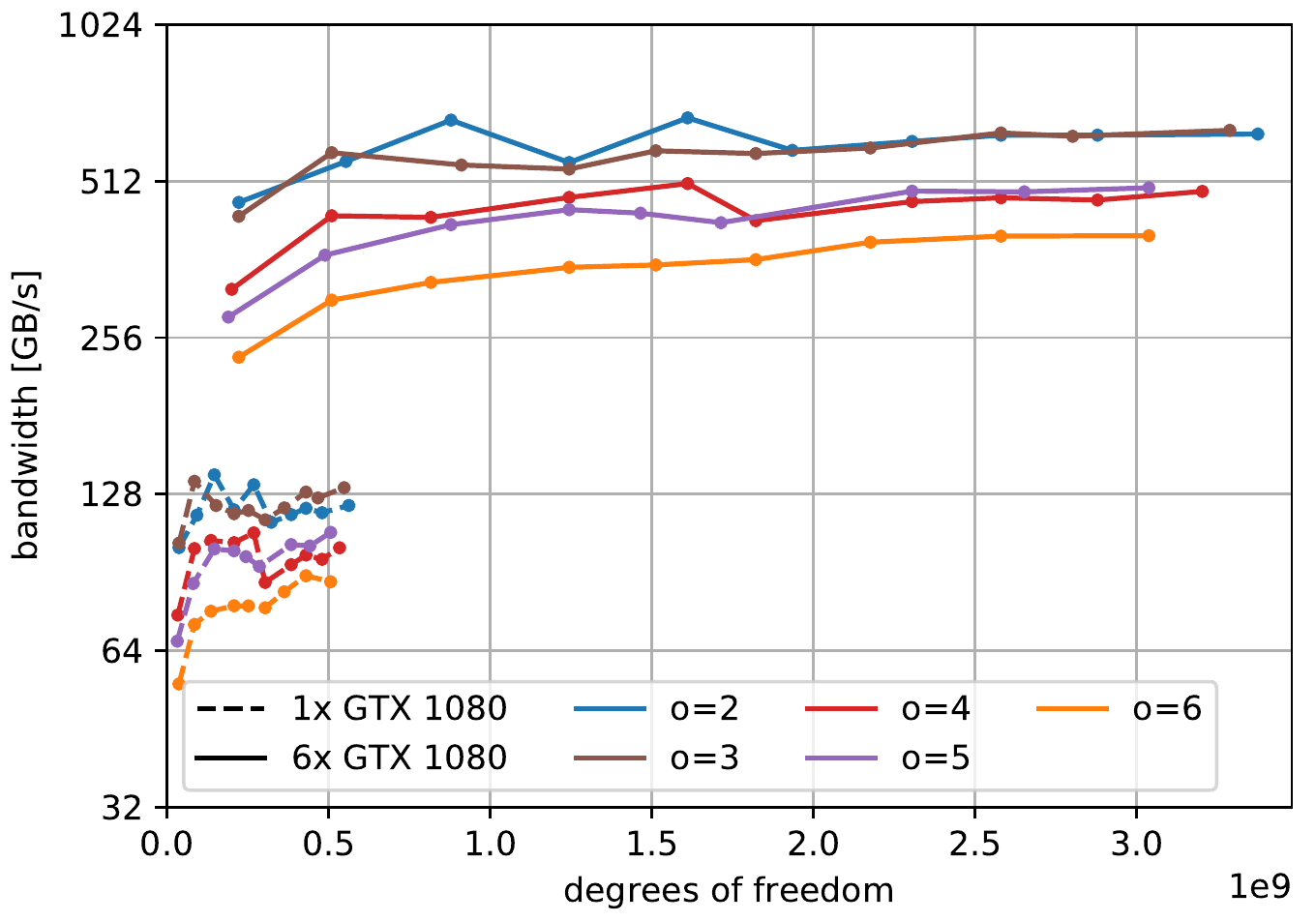}\hspace{1cm}\includegraphics[width=7cm]{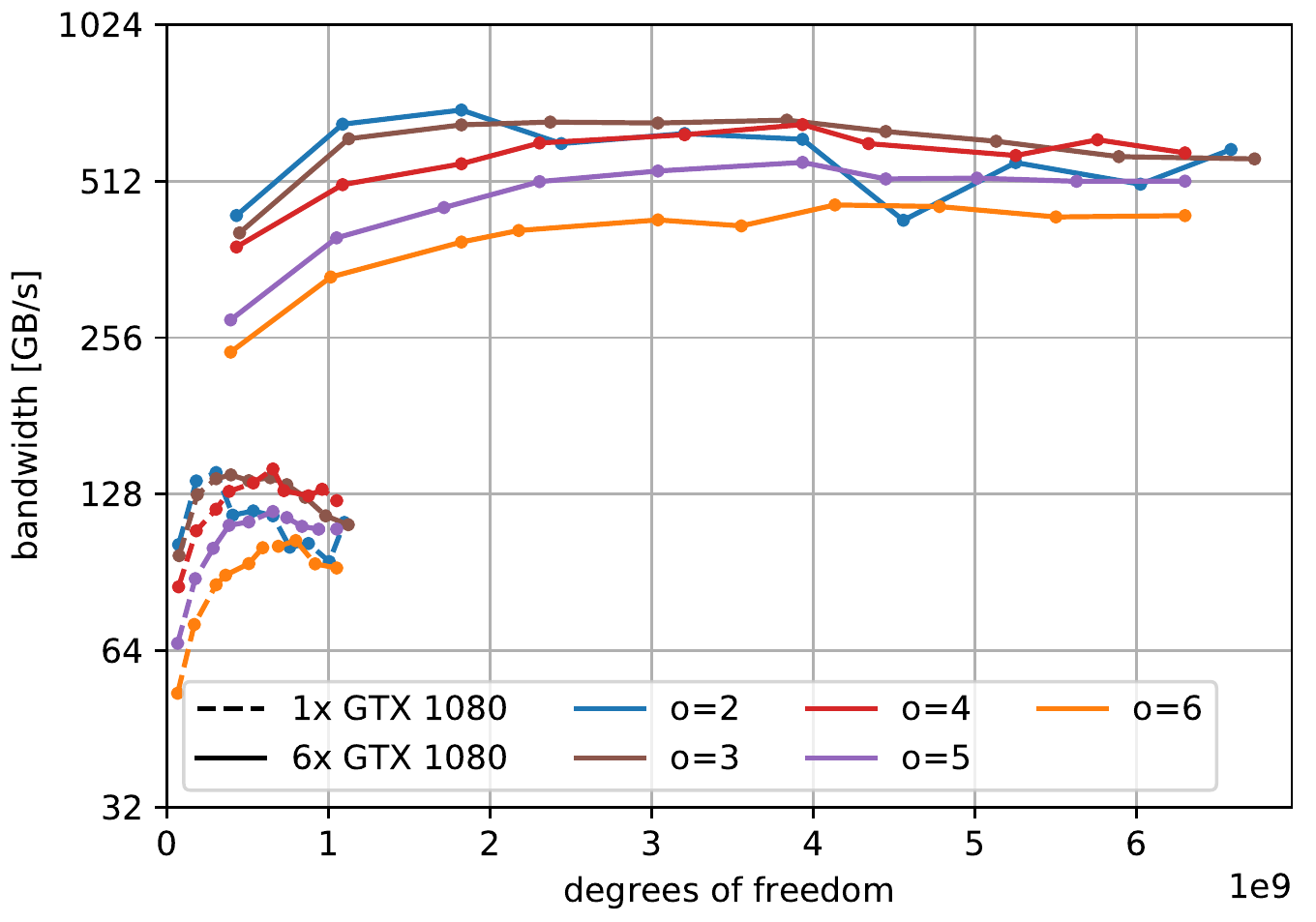}
\par\end{centering}
\caption{Performance of six GTX 1080 Ti GPUs on a single node for solving the
2+2 dimensional Vlasov--Poisson equation. The achieved bandwidth
is shown as a function of the number of degrees of freedom. The simulation
are performed using double precision (left) and single precision (right)
floating point numbers. For comparison the results for a single GPU
of the same type are shown as well.\label{fig:gtx1080}}
\end{figure}

The goal of the present section is to investigate the performance
of such a consumer level GPU for our application. As an example, we
have chosen the GTX 1080 Ti; see Table \ref{tab:hardware-specs} for
more details. The corresponding results are shown in Figure \ref{fig:gtx1080}.
We observe that the difference in performance for this GPU between
single and double precision performance is not as large as one might
expect. However, remember that we have a memory bound problem and
thus the inability of the GTX 1080 Ti to do double precision operations
natively has not a drastic effect on performance. The difference for
the 4th order scheme, for example, is approximately 30\% in terms
of efficiency (i.e.~approximately a factor of $2.6$ between single
and double precision simulations). Note, however, that even in the
single precision case the GTX 1080 Ti achieves only a fraction of
the performance compared to the V100 or Titan V (132 GB/s vs 411 GB/s
and 350 GB/s, respectively, for the fourth order method). In the multiple
GPU setting, 6 GTX 1080 Ti are able to achieve a combined single precision
bandwidth of 612 GB/s. This is just short of twice the performance
of a Titan V. 

\begin{figure}[H]
\begin{centering}
\includegraphics[width=7cm]{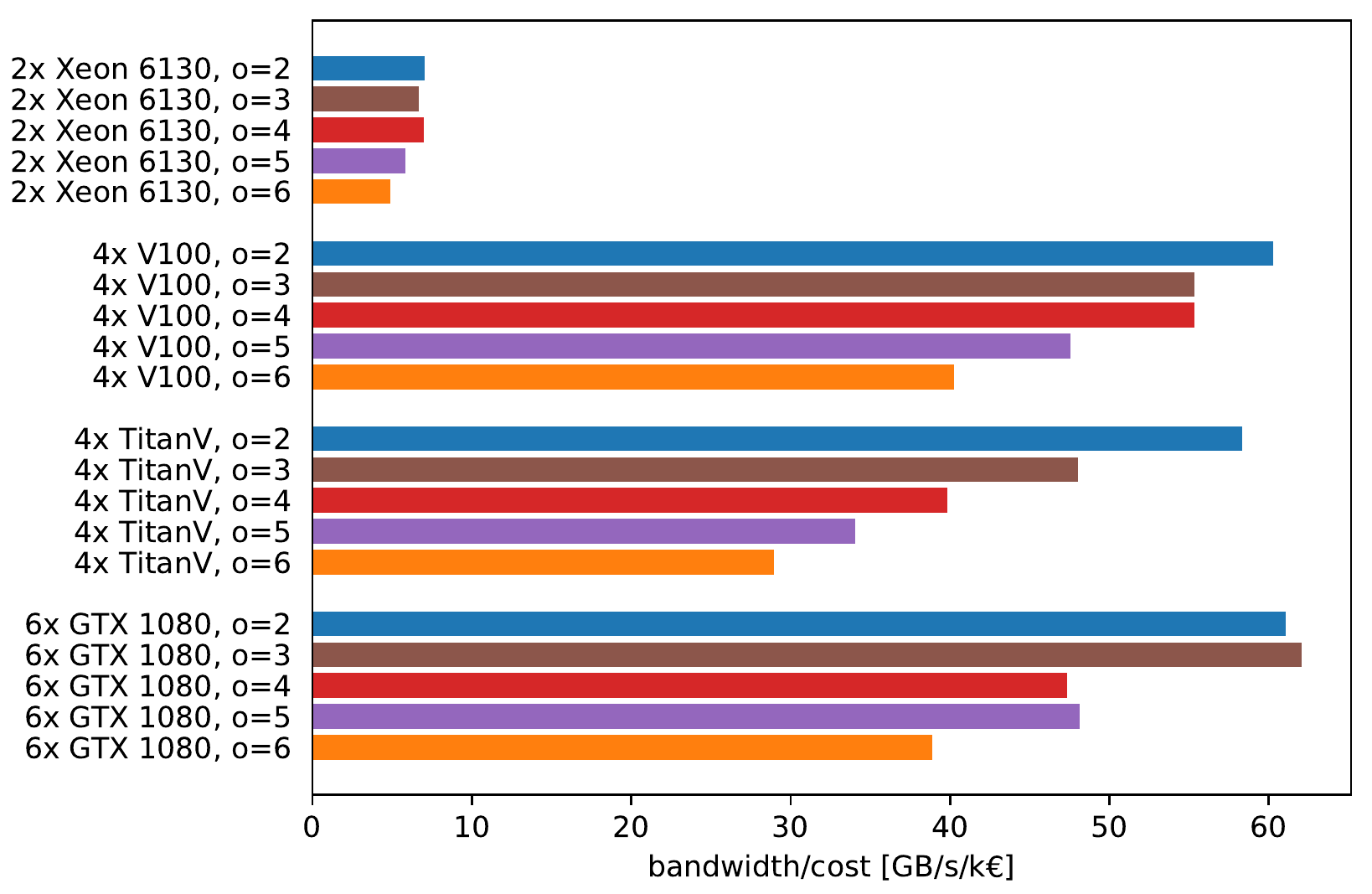}\hspace{1cm}\includegraphics[width=7cm]{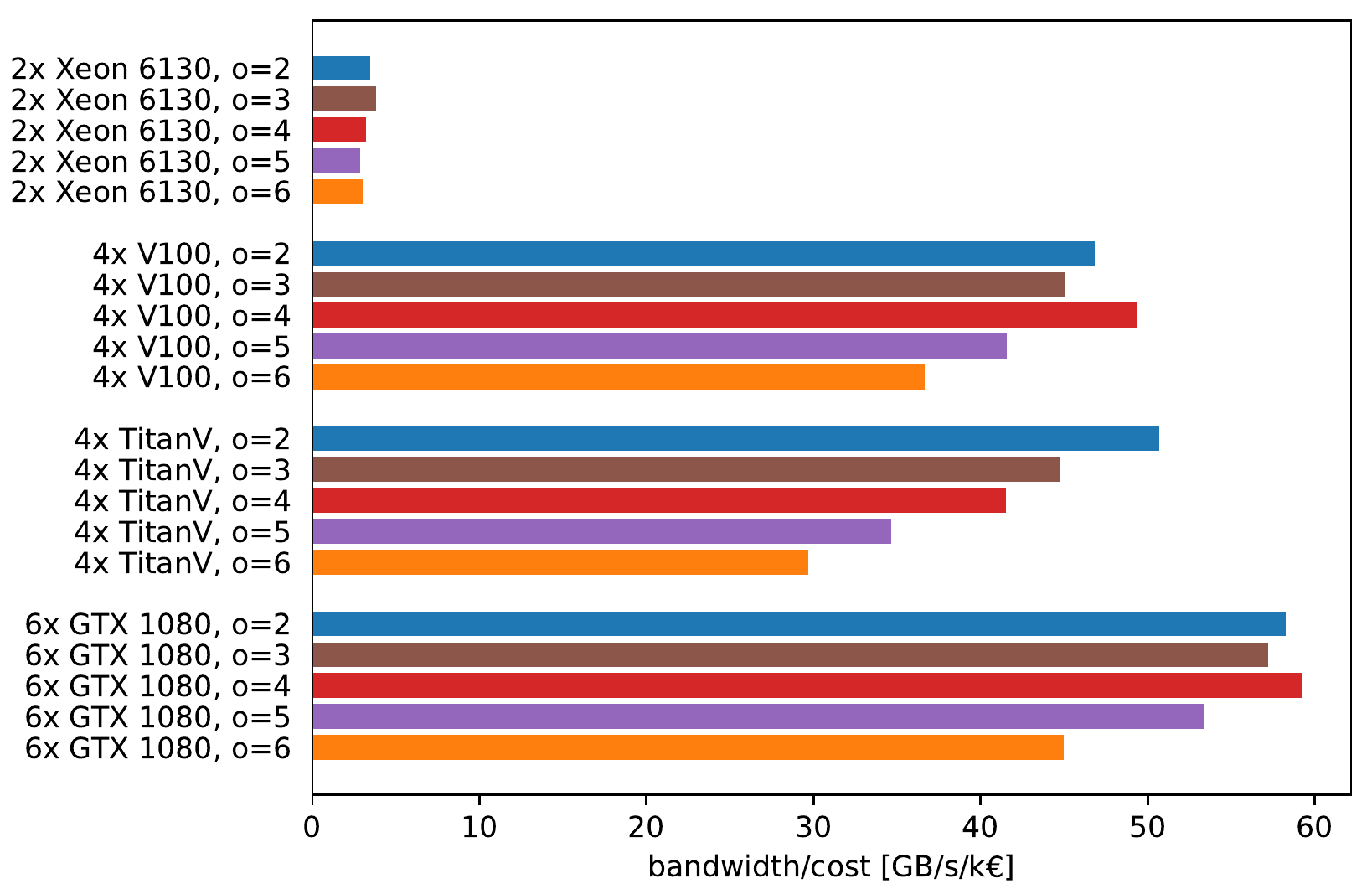}
\par\end{centering}
\caption{\textcolor{black}{Performance normalized to cost for solving the 2+2
dimensional Vlasov--Poisson equation. The achieved bandwidth per
1000 € of procurement cost on a single node is shown. The simulation
are performed using double precision (left) and single precision (right)
floating point numbers. \label{fig:performance-per-cost}}}
\end{figure}

\textcolor{black}{In Figure \ref{fig:performance-per-cost} we investigate
the performance normalized to procurement cost. We see that for both
double and single precision the GTX 1080 Ti cards have an advantage
here. Despite this, the V100 and TitanV perform very similar according
to this this metric. The overall advantage of any GPU solution compared
to the dual socket Xeon Gold 6130 node is approximately a factor of
$6$ to $7$.}

\section{Performance in 1+3d and 2+3d\label{sec:1x3v-and-2x3v}}

As described in the introduction, our implementation is able to handle
arbitrary dimensions, in both the $x$ and $v$-direction, within
the same code base. This is, in particular, an advantage if good performance
for different configurations can be achieved. In this case a single
code base is sufficient to treat all the different configurations
and no additional code to specifically optimize certain configurations
has to be written. This simplifies the maintenance of the software
package enormously. To demonstrate that this is indeed the case for
SLDG we will consider 1+3 dimensional (i.e.~1 dimension of space
and 3 dimensions of velocity) and 2+3 dimensional (i.e.~2 dimensions
in space and 3 dimensions in velocity) simulation in this section.
For brevity, we will only present results for double precision computations
and V100 GPUs here. However, we emphasize that there are no surprises
when looking at the performance of the other GPUs considered in this
paper.

\begin{figure}[H]
\begin{centering}
\includegraphics[width=7cm]{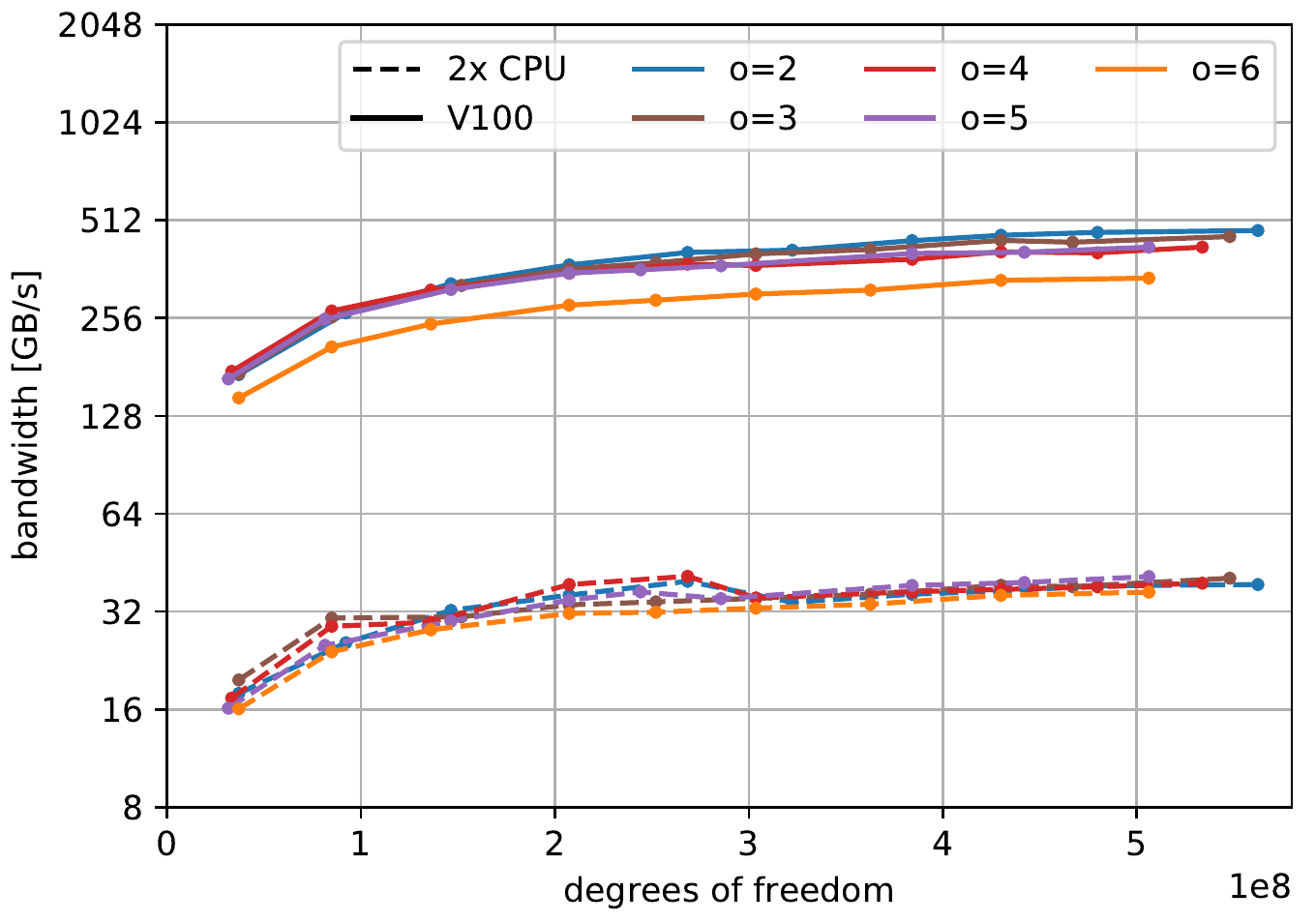}\hspace{1cm}\includegraphics[width=7cm]{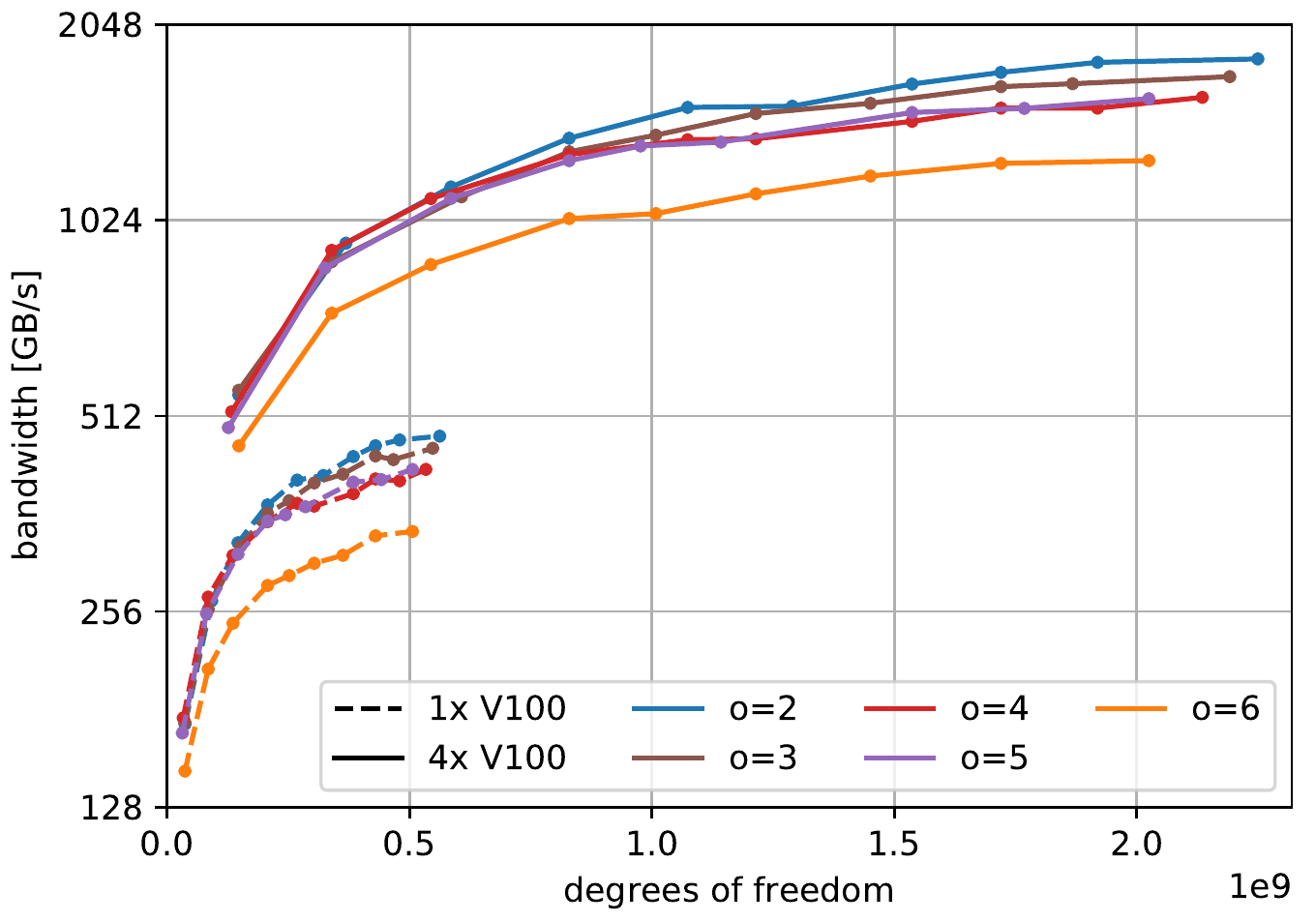}
\par\end{centering}
\caption{Performance of V100 GPUs for solving the 1+3 dimensional Vlasov--Poisson
equation. The achieved bandwidth for a single V100 (left) and four
V100 (right) is shown as a function of the number of degrees of freedom.
The simulation are performed using double precision floating point
numbers. \label{fig:1x3v-nvl}}
\end{figure}

Let us start with the 1+3 dimensional case. The corresponding results
for the configuration from sections \ref{sec:single-gpu} and \ref{sec:multiple-gpu}
are shown in Figure \ref{fig:1x3v-nvl}. In this setting the performance
is very similar to the 2+2 dimensional case considered so far. For
example, for the fourth order scheme we achieve a bandwidth of approximately
424 GB/s for a single V100 and 1584 for four V100 connected via NVLink.
We note that the difference between the schemes with different order
is less pronounced; even the sixth order scheme achieves 1265 GB/s
on four V100 GPUs. Before proceeding let us note that the performance
on the CPU is very similar to what has been observed in section \ref{sec:single-gpu}
for the 2+2 dimensional case. We achieve approximately 40 GB/s. Also
in this case the spread between the schemes of different order is
significantly reduced.

\begin{figure}[H]
\begin{centering}
\includegraphics[width=7cm]{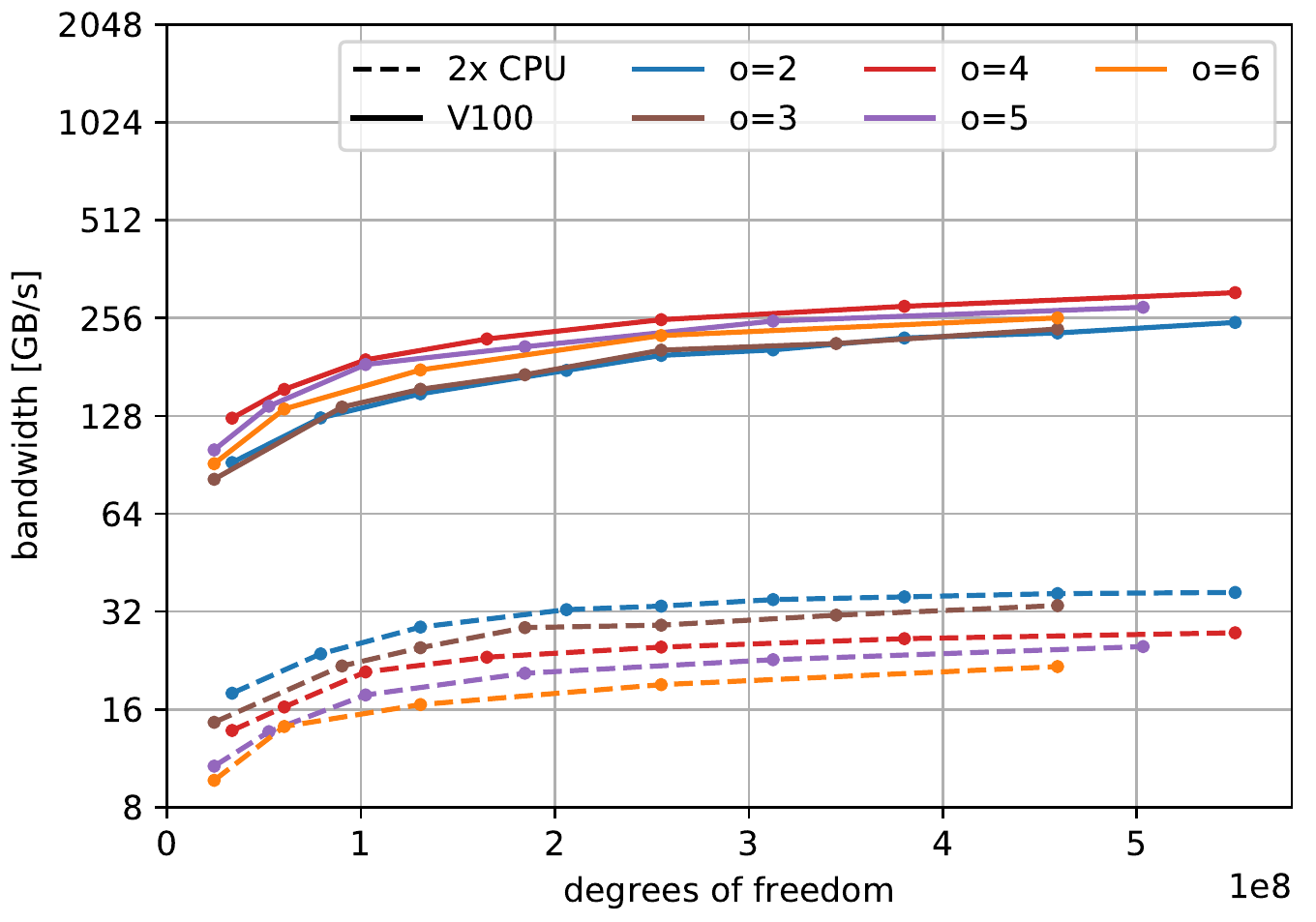}\hspace{1cm}\includegraphics[width=7cm]{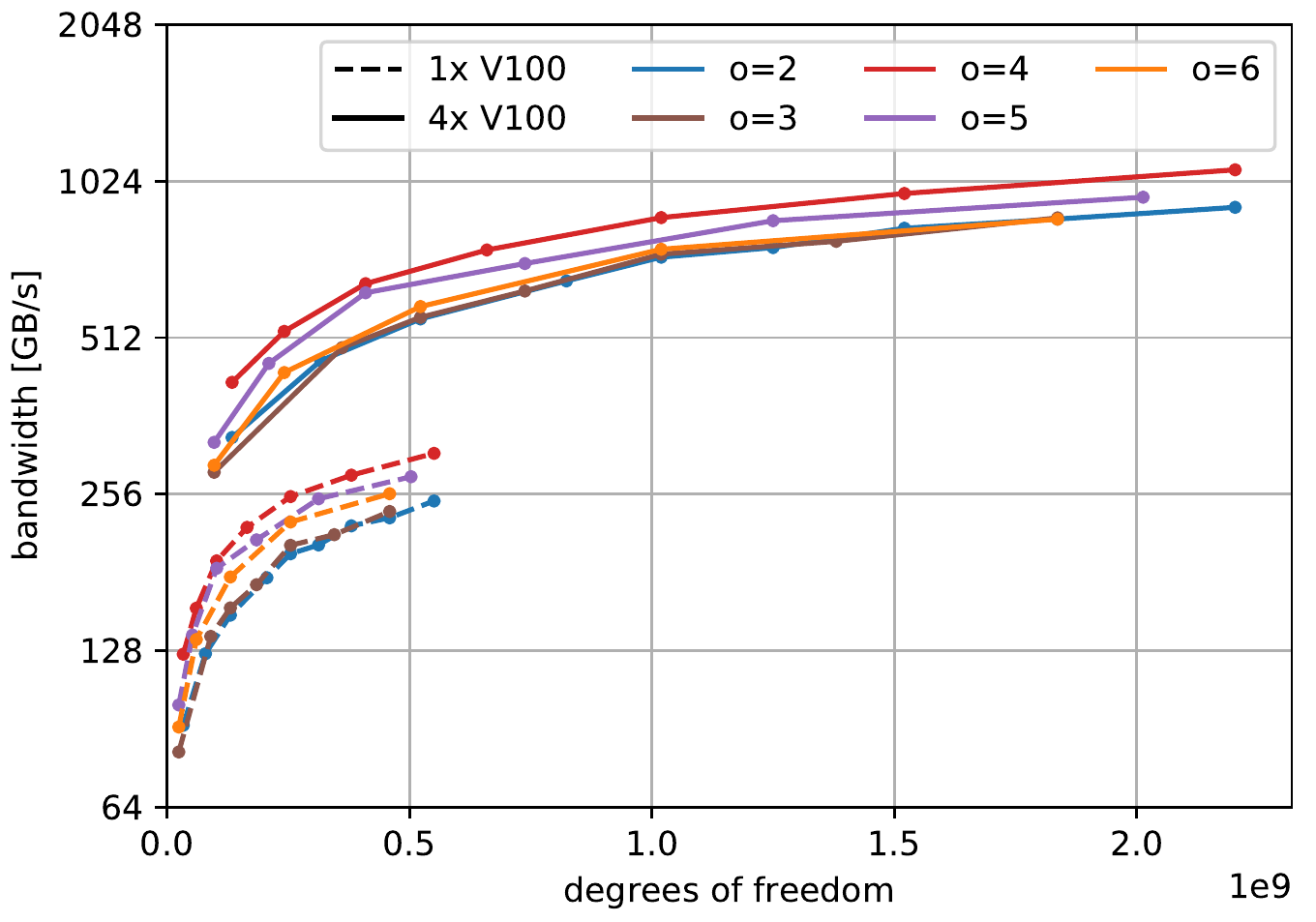}
\par\end{centering}
\caption{Performance of V100 GPUs for solving the 2+3 dimensional Vlasov--Poisson
equation. The achieved bandwidth for a single V100 (left) and four
V100 (right) is shown as a function of the number of degrees of freedom.
The simulation are performed using double precision floating point
numbers.\label{fig:2x3v-nvl}}
\end{figure}

We now consider the 2+3 dimensional case. Thus, we simulate a five
dimensional problem. The obtained results are shown in Figure \ref{fig:2x3v-nvl}.
We remark that the achieved bandwidth is slightly reduced compared
to the 2+2 dimensional case. For example, for the fourth order method
a single V100 achieves 307 GB/s and four V100 achieve 1077 GB/s. A
similar reduction in performance can be observed for the CPU and thus,
overall, the difference between the CPU and the single GPU implementation
is still approximately a factor of $10$.

\section{Numerical simulation \& validation\label{sec:numerical-simulation}}

In this section we will discuss the steps we have taken to validate
our code. In SLDG we make use of a range of automated tests. This
includes unit testing the components \textcolor{black}{and algorithms
}that make up the software package (using \texttt{Boost.Test}).\textcolor{black}{{}
In particular, automatic tests check the convergence rate of all numerical
algorithms involved and test their accuracy, in cases where analytic
solutions can be constructed. In addition, we test} the results of
numerical simulations for the complete Vlasov equation\textcolor{black}{.}
For the latter we make use of analytic properties that are available
in certain situations. For example, we have a test where a multi-dimensional
linear Landau damping simulation is conducted. Since the analytic
decay rate (as a function of the wave number of the initial perturbation)
is known, we use this to check the output of the numerical simulation.
We also use properties of the numerical algorithm, such as mass and
momentum conservation, to further verify the code. For all the numerical
simulations that have been conducted in order to obtain the results
presented in sections \ref{sec:single-gpu}-\ref{sec:1x3v-and-2x3v}
we automatically check that all runs give the same results (within
the approximation error).

\begin{figure}[h]
\centering{}\includegraphics[width=16cm]{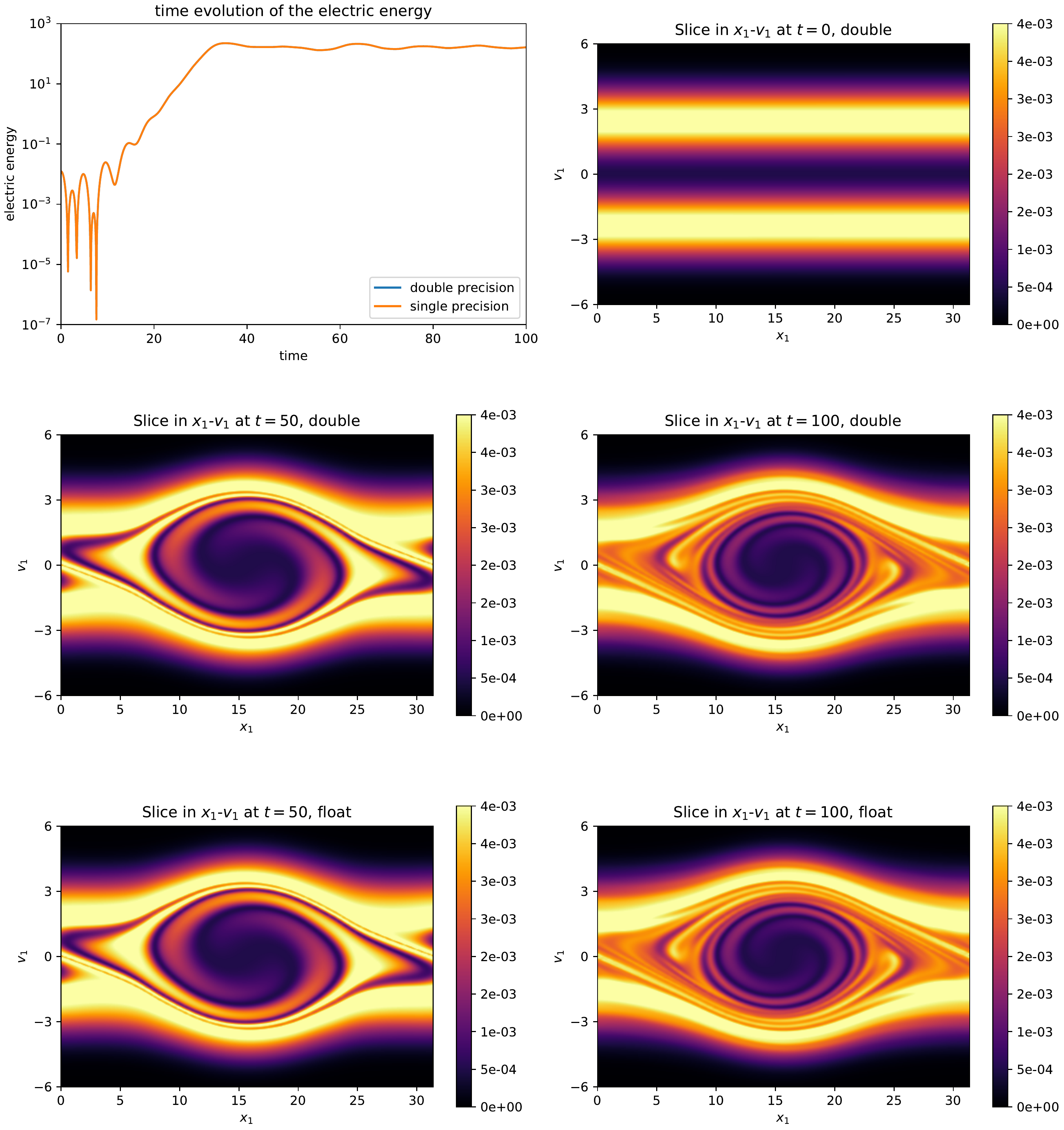}\caption{Numerical simulation of the two-stream instability given by equation
(\ref{eq:iv-tsi}). On the top-left the time evolution of the electric
field is shown (note that the double and single precision results
almost overlap). The remaining plots are $x_{1}-v_{1}$ slices of
the density function $f$ at time $t=0$ (top-right), $t=50$ (middle-left
for double precision and bottom-left for the single precision simulation),
and $t=100$ (middle-right for double precision and bottom-right for
the single precision simulation). The simulation is conducted on a
grid of size $60^{4}\cdot4^{4}$. That is, in each coordinate direction
$60$ cells, each with $o=4$ degrees of freedom (i.e.~the fourth
order semi-Lagrangian discontinuous Galerkin scheme) are used. \label{fig:tsi}}
\end{figure}

\begin{figure}[h]
\centering{}\includegraphics[width=7cm]{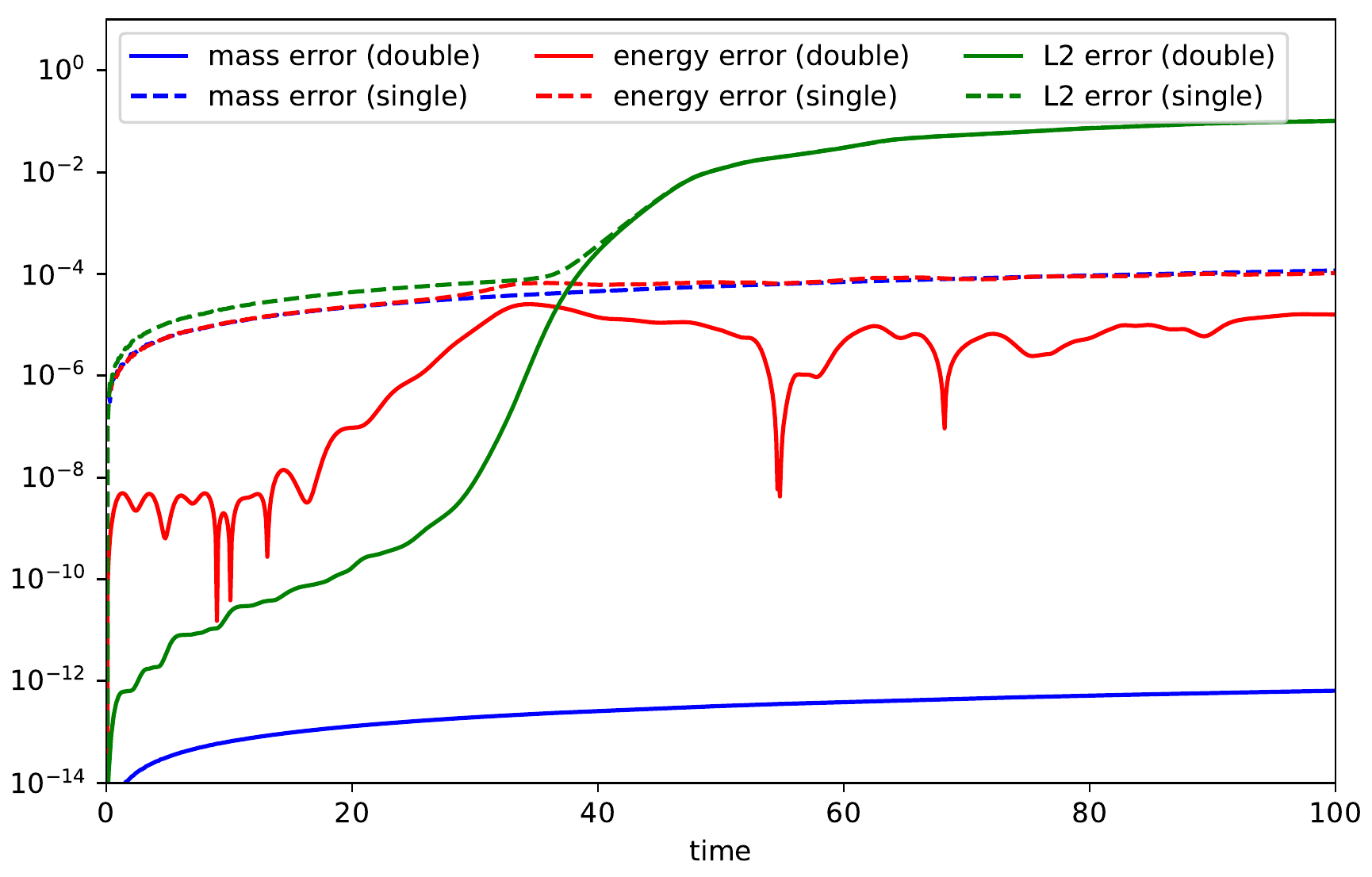}\caption{\textcolor{black}{Numerical simulation of the two-stream instability
given by equation (\ref{eq:iv-tsi}). The time evolution of the relative
error in mass, energy, and $L^{2}$ norm is shown. The simulation
is conducted on a grid of size $60^{4}\cdot4^{4}$. That is, in each
coordinate direction $60$ cells, each with $o=4$ degrees of freedom
(i.e.~the fourth order semi-Lagrangian discontinuous Galerkin scheme)
are used. \label{fig:tsi-difference}}}
\end{figure}

Finally, we compare the output of numerical simulations, for which
no analytic properties are available, to known results from the literature.
We will present one such simulation here. Namely, a four-dimensional
two-stream instability given by the initial data
\begin{align}
f(0,x_{1},x_{2},v_{1},v_{2}) & =(1+\epsilon(\cos(kx_{1})+\cos(kx_{2})))f^{\text{eq}}(v_{1},v_{2})\label{eq:iv-tsi}
\end{align}
with 
\[
f^{\text{eq}}(v_{1},v_{2})=\frac{1}{8\pi}\left(\mathrm{e}^{-(v_{1}-v_{0})^{2}/2}+\mathrm{e}^{-(v_{1}+v_{0})^{2}/2}\right)\left(\mathrm{e}^{-(v_{2}-v_{0})^{2}/2}+\mathrm{e}^{-(v_{2}+v_{0})^{2}/2}\right),
\]
where $\epsilon=10^{-3}$, $k=0.2$, and $v_{0}=2.4$. The computational
domain is $[0,10\pi]^{2}\times[-6,6]^{2}$. The time evolution of
the electric field and some snapshots of the density function are
shown in Figure \ref{fig:tsi}. We remark that these results agree
well with what has been reported in the literature (see, for example,
\cite{Kormann15,barsamian2018verification}). Figure \ref{fig:tsi}
shows both single and double precision results, which are almost indistinguishable
in the plots. \textcolor{black}{The difference in the distribution
function (measured in the infinity norm), the electric field (measured
in the infinity norm), and the relative difference in electric energy
are below $2\cdot10^{-4}$. In addition, we show the time evolution
of the physical invariants mass, energy, and $L^{2}$ norm in Figure
\ref{fig:tsi-difference}. We observe that for both energy as well
as the $L^{2}$ norm the difference between double and single precision
result is very small.}

\section{Conclusion\label{sec:conclusion}}

We have shown that the dimension independent semi-Lagrangian Vlasov
solver SLDG can achieve excellent performance on systems with single
and multiple GPUs. In fact, the single node GPU performance (i.e.~4
V100 GPUs on the same node compared to the dual socket CPU system)
shows a speed up of approximately a factor of $35$ compared to a
dual socket CPU node. We further emphasize that the Titan V also shows
excellent performance characteristics. The main performance bottleneck
for the Titan V in the multi-GPU setup is the absence of a fast interconnect
(such as NVLink).

Although there are some variations, the observed performance is rather
predictable across 2+2 dimensional, 1+3 dimensional, and 2+3 dimensional
Vlasov simulations. This is a good validation of our approach to develop
a dimension independent code. It is also interesting to note that
higher order methods are less efficient than lower order methods.
One might be tempted to blame this on the increased computational
burden. However, as discussed, the increased data transfer plays an
even larger role. This is particularly true for GPUs without a fast
interconnect. Performance on the GTX 1080 Ti is rather disappointing
giving the theoretical memory bandwidth of those GPUs. However, on
a pure performance per cost basis consumer cards are still winning
out. It is interesting to note that this is true for both single and
double precision computations. That is, there is only a modest performance
penalty for the double precision implementation on the GTX 1080 Ti.

\textcolor{black}{The current version of the SLDG code can be found
at \url{https://bitbucket.org/leinkemmer/sldg}. The code can be freely
used under the terms of the MIT license.}

\bibliographystyle{plain}
\bibliography{sldg-gpu}

\end{document}